\documentclass[superscriptaddress, 10pt, aps, pra, twocolumn, showpacs, longbibliography, showkeys]{revtex4-2}
\usepackage{amsmath,amssymb,amsthm}
\usepackage{easybmat}
\usepackage[colorlinks=true,citecolor=blue,urlcolor=blue, linkcolor= magenta]{hyperref}
\usepackage[pdftex]{graphicx}
\usepackage{times, txfonts}
\usepackage{braket}
\usepackage{color}
\usepackage{float}
\usepackage{bbold}
\newcommand{\be}{\begin{equation}}
	\newcommand{\ee}{\end{equation}}
\newcommand{\ba}{\begin{eqnarray}}
	\newcommand{\ea}{\end{eqnarray}}
\newcommand{\ketbra}[2]{|#1\rangle \langle #2|}

\newcommand{\Tr}{{\rm Tr}}

\begin{document}
	
	\title{Collective Quantum Batteries and Charger-Battery Setup in Open Quantum Systems: Impact of Inter-Qubit Interactions, Dissipation, and Quantum Criticality}

	\author{Mahima Yadav\textsuperscript{}}
	\email{yadav.15@iitj.ac.in}
	\author{Devvrat Tiwari\textsuperscript{}}
	\email{devvrat.1@iitj.ac.in }
    \author{Subhashish Banerjee\textsuperscript{}}
	\email{subhashish@iitj.ac.in }
	\affiliation{Indian Institute of Technology Jodhpur-342030, India\textsuperscript{}}

\date{\today}

\begin{abstract}
Quantum batteries have emerged as promising platforms for exploring energy storage and transfer processes governed by quantum mechanical laws. In this work, we study three models of two-qubit open quantum systems. The first model comprises two central spins immersed in spin baths, and both central spins are collectively considered as quantum batteries. The impact of inter-qubit interactions on the performance of the quantum battery is investigated. In the second model, a two-qubit model interacting with a squeezed thermal bath serves as a collective quantum battery, where the impact of inter-atomic distance and the bath temperature on the battery's performance is explored. Furthermore, a two-qubit model is used, where one qubit is modeled as a battery and the other as a charger. The charger in this model interacts with an anisotropic spin-chain bath, which is conducive to quantum criticality. It is demonstrated that this criticality has a substantial impact on the quantum battery's storage capacity. 
\end{abstract}

\maketitle

\section{Introduction}\label{sec-intro}

Quantum thermodynamics provides a framework for understanding how the foundational laws of thermodynamics apply to quantum systems~\cite{gemmer2004quantum, binder2019thermodynamics, sai_anders_book, deffner2019quantum, Seifert_2012, Hanggi_talkner, sekimoto2010stochastic, Alicki2018_Kosloff}. One of the primary objectives of this field is to correctly define the laws of thermodynamics governing energy exchange and entropy production in the quantum regime and understand the thermalization of quantum systems~\cite{landi_review_entropy_production, devvrat_strong, SB2023_thermalization, devvrat-neha_hmf}. Recent advances have shown that exclusive quantum phenomena, such as entanglement and coherence, can act as thermodynamic resources~\cite{binder2019thermodynamics, Gour_2022}.
With the advancement in quantum technologies, quantum thermodynamics has been utilized in the development of various quantum thermal devices, ranging from heat engines~\cite{Alicki_1979, thomas_heat_engine, KUMAR2023128832, prasanna_heat_engine}, thermal analogs of electric devices~\cite{devvrat_circuit}, such as diodes~\cite{diode_2017}, transistors~\cite{transistor_2016, floquet_transistor_2022}, adders~\cite{devvrat_circuit}, and Wheatstone bridges~\cite{wheatstone_bridge_2022, devvrat_circuit}, and quantum batteries~\cite{QB_colloquium, QB_andolina_2019, QB_open_system, Devvrat_impact, bhanja2023, yadav2025thermo}, among others. Quantum batteries have recently garnered significant attention, demonstrating a quantum advantage in energy storage and transfer~\cite{QB_advantage_1, Shastri2025, QB_advantage_3}. 

Quantum batteries are quantum mechanical systems designed to store energy temporarily for future use~\cite{Alicki_Fannes_QB, Binder_2015, QB_colloquium, campaioli2018quantum}. The quantum battery can leverage the properties of a quantum system, such as coherence and entanglement, to gain an advantage in charging and power delivery rates~\cite{Campaioli_2017, Uwe_Fischer_QB}. Quantum batteries have been realized on a number of platforms, including the Dicke model quantum battery with its extended Dicke model variant~\cite{Ferraro_2018, Extended_Dicke_battery}, spin-chain quantum batteries~\cite{spin_chain_quantum_battery, central_spin_quantum_battery}, strongly interacting Sachdev-Ye-Kitaev fermionic battery~\cite{SYK_battery}, solid-state quantum battery~\cite{Ferraro_2018}, self-discharge-mitigated quantum battery~\cite{self_discharge_mitigated_QB}, resonator-qutrit quantum battery~\cite{resonator-qutrit_QB}, Rosen-Zener quantum battery~\cite{Rosen-Zener_QB}, topological quantum batteries~\cite{topological_QBs}, and Unruh-DeWitt battery~\cite{mukherjee2024_QB}, among others. The maximum amount of work that can be extracted from a quantum battery is quantified by ergotropy~\cite{Allahverdyan_2004}. In finite quantum systems, the passive state of a system (the state that can not perform work) is usually different from the Gibbs state and depends on the spectral decompositions of the state of the system and its Hamiltonian~\cite{Pusz1978}. This can be further divided into its coherent and incoherent parts to accommodate the impact of population and coherence of a quantum state on ergotropy~\cite{coherent_ergo1}. Further, the charging power is used to characterize the charging and discharging process of a quantum battery~\cite{Devvrat_impact}. In realistic scenarios, the quantum battery--a quantum mechanical system--can not remain perfectly isolated from its ambient environment, and the environment has a significant impact on the work storage capacity of a quantum battery~\cite{cakmak1, Kamin_2020, bhanja2023, yadav2025thermo, Devvrat_impact, rishav_QB, malavazi2025_QB, AhmadiB_2024nonreciprocal, ahmadi2025harnessing, Ahmadi_2025superoptimal}. 

The theory of open quantum systems accounts for the environmental effects on a quantum system~\cite{breuer2002book, Weiss, sbbook}. Traditionally, the evolution of open quantum systems has been examined using the Markovian Gorini-Kossakowski-Sudarshan-Lindblad (GKSL) master equation~\cite{gksl_master1, Lindblad1976}. However, recently, rapid inroads have been made in the challenging domain of non-Markovian dynamics. The presence of non-Markovian evolution has been extensively explored and applied to various problems~\cite{de_vega_alonso, LI20181, Cresser_2014, Rivas_2014, CHRUSCINSKI20221, Utagi2020, kading2025}. From the perspective of a quantum battery, non-Markovian evolution has been seen as a recharging mechanism, where the environment recharges the system after an initial discharge~\cite{Kamin_2020, Morrone_2023, bhanja2023, Devvrat_impact}. 

The reservoir interacting with an open quantum system can be broadly classified into two types: a bosonic bath and a spin bath. The bosonic bath has been prototypical for a wide variety of open quantum systems, including the Caldeira-Leggett model and the spin-boson model, among others~\cite{CALDEIRA1983_tunneling, Legget_dissipative}. The spin baths, composed of a finite number of spins, were initially observed in magnetic systems~\cite{prokofev_stamp_2000, Prokofaev1996}. Thereafter, this has been utilized in several systems and has been experimentally simulated using Rydberg atoms, quantum dots, and NV-centers, and has been incorporated into the experimental realization of a quantum battery~\cite{Haase_2018, Hanson_2006, trapped_ions, TSMahesh_2022}. The central spin model serves as the primary framework for studying spin baths~\cite{Burgarth2004, Bhattacharya2021, Chiranjib_2017, PhysRevA.106.032435, devvrat_central_spin_2, devvrat_strong}. In a recent study, it has been demonstrated that replacing the non-interacting spin-bath with an anisotropic spin-chain bath in a central spin model results in distinct critical behavior, leading to a quantum phase transition~\cite{Wei2016}. 

Quantum phase transitions, driven by quantum fluctuations, have drawn considerable interest~\cite{sachdev2011quantum, Preskill2000, Dutta2015}. Quantities such as entanglement, quantum speed limit time, and non-Markovianity have been found to be significantly impacted by these transitions~\cite{Vidal_QPT, Osterloh2002, Zhang_2007}. It is imperative to study the impact of quantum critical behavior on the performance of a quantum battery.  
 
In this work, we take up three two-qubit open quantum system models, two of which are envisaged as collective quantum batteries, and a charger-battery setup is modeled by the last one. The effects of inter-qubit interactions, dissipation by the bath, and quantum criticality on the performance of these quantum batteries are studied. Recently, quantum batteries have been realized with Dzyaloshinskii–Moriya (DM) interaction~\cite{Moriya, DZYALOSHINSKY1958} between the qubits~\cite{DM_battery_2, DM_battery, bhattacharya2025, vigneshwar2026noise}. Dipolar spin systems with coexisting dipole–dipole and DM interactions have also been explored to enhance quantum battery performance by exploiting quantum coherence~\cite{parkavi2026tunable}. This motivates us to compare the impact of inter-qubit interaction on a quantum battery. This is addressed using a two-qubit central spin model interacting with spin-baths~\cite{PhysRevA.106.032435, devvrat_central_spin_2}, where the two qubits are collectively considered as a quantum battery, and the impact of XXX~\cite{Heisenberg1928, Bethe1931, Mattis1981, spin_chain_primer_nepomechie} and DM inter-qubit interactions on its performance is compared. Furthermore, to explore the bath-assisted dissipation of the quantum battery, a two-qubit collective decoherence model interacting with a squeezed thermal bath is used. This model has been used to demonstrate entanglement generation between the qubits assisted by the bath, depending on the inter-qubit distance~\cite{Banerjee_2007, BANERJEE2010, FICEK2002}. We investigate how this inter-qubit distance and the temperature of the bath affect the ergotropy of the collective two-qubit quantum battery. In a recent work, ergotropy and entanglement in interacting spin systems have been studied in the context of critical spin chains, illustrating how entanglement and system size affect extractable work in many-body spin models~\cite{MulaB_2023}. In this spirit, we explore the impact of quantum critical behavior on the dynamics of the quantum battery  using a novel two-qubit central spin charger-battery setup. In this model, the charger qubit interacts with an anisotropic Heisenberg XY spin chain, and the battery is immersed in a non-interacting spin bath. This model demonstrates that criticality has a significant effect on the storage of charge in a quantum battery. 

The plan of the paper is as follows. In Sec.~\ref {sec_prelim}, we discuss the quantifiers used to investigate the performance of the battery and the models envisioned as a quantum battery. Section~\ref{sec_collective_battery} discusses the impact of inter-qubit interactions on the collective central spin battery and inter-atomic distance, as well as bath temperature, on the two-qubit collective decoherence quantum battery. In Sec.~\ref{sec_charger_battery_setup_investigation}, the central spin charger-battery setup is discussed, and the impact of quantum criticality on the battery's performance is investigated. The conclusions are presented in Sec.~\ref{conclusion}. 

\section{Preliminaries} \label{sec_prelim}
\subsection{Characterizers of a quantum battery}
To investigate the performance of a quantum battery, we use quantifiers such as ergotropy, energy, instantaneous, and average (dis-)charging powers, which are briefly discussed in this section.
\subsubsection{Ergotropy}
The ergotropy of a quantum system of dimension $d$, with the system Hamiltonian $H_S$ and state $\rho_S(t)$ at any time $t$, is given by~\cite{Allahverdyan_2004}
\begin{align}\label{eq_ergotropy}
\mathcal{W}[\rho_S (t)] = \text{Tr} \left[\rho_S (t) H_S \right] - \text{Tr} \left[\rho_S^p(t) H_S \right],
\end{align}
where \( \rho_S^p(t) \) is the passive state corresponding to the input state \( \rho_S(t) \), from which no additional work can be extracted. The passive state is reached when the system undergoes an optimal unitary evolution driven by a cyclic potential. In contrast to macroscopic systems, where passive states typically coincide with thermal Gibbs states, the passive state of a finite-dimensional quantum system is in general distinct from a Gibbs state.  
The passive state for a microscopic quantum system is given by
\begin{align}
    \rho_S^p(t) = \sum_{j=1}^{d} r_j |\epsilon_j\rangle \langle \epsilon_j|,
\end{align}
where the $\ket{\epsilon_j}$'s come from the spectral decomposition of the system's Hamiltonian
$H_S = \sum_j \epsilon_j |\epsilon_j\rangle \langle \epsilon_j|$. The spectral decomposition of $\rho_S(t) = \sum_j r_j \ket{r_j}\bra{r_j}$ is used to get the $r_j$'s in the above equation. Importantly, the eigenvalues \( r_j \) and \( \epsilon_j \) follow the order
\begin{equation}
    r_1 \geq r_2 \geq r_3 \dots \geq r_d, \quad \text{and} \quad \epsilon_1 \leq \epsilon_2 \leq \epsilon_3 \dots \leq \epsilon_d\, .
    \label{eq_spectral_order}
\end{equation}
The above ordering ensures that the lowest-energy state has the highest population, and as the level's energy increases, its population decreases, making it passive with respect to work extraction.

The ergotropy can further be divided into coherent $\mathcal{W}_{c}$ and incoherent ergotropy $\mathcal{W}_{i}$, such that $\mathcal{W} = \mathcal{W}_{i} + \mathcal{W}_{c}$ accommodating contributions of coherence and population terms of the system's state~\cite{coherent_ergo1}. The incoherent ergotropy is the amount of work that can be extracted from the system without altering its coherence by utilizing the dephased state of the system, and is given by
\begin{align}\label{eq_incoherent_ergotropy}
\mathcal{W}_{i}\left[\rho_S(t)\right] = {\rm Tr}\left[\left\{\rho^{D}_S(t)- \rho^{D}_{p}(t)\right\} H_S\right], 
\end{align}
where $\rho^{D}_S(t) = \sum_{i} \bra{i}\rho_S(t) \ket{i} \ketbra{i}{i} $ denotes the dephased state for the system state $\rho_S(t)$, and $\rho^{D}_{p}(t)$ is the passive state corresponding to $\rho^{D}_S(t)$. The coherent ergotropy is then readily obtained as $\mathcal{W}_{c}[\rho_S(t)] = \mathcal{W}[\rho_S(t)] - \mathcal{W}_{i}[\rho_S(t)]$.
\subsubsection{Energy and power}
The energy of the system at any time $t$, using the Hamiltonian $H_S$ and state of the system $\rho_S(t)$, is given by 
\begin{align}
E(t) = {\rm Tr}\left[H_S\rho_S(t)\right].
\end{align}
The corresponding instantaneous power of the system is defined as the time derivative of the energy, and is given by
\begin{align}
    P(t) = \frac{dE(t)}{dt}.
    \label{eq_inst_power}
\end{align}
The energy and the power quantify the stored energy and its variation in the quantum battery. However, the amount of work that can be extracted from the quantum battery is quantified by ergotropy.  

\subsubsection{Charging power}
Ergotropy is used to characterize the charging and discharging behavior of the quantum battery. The quantum battery is said to charge when ergotropy increases and to discharge when ergotropy decreases. The time derivative of ergotropy is the charging power given by
\begin{align}\label{eq_charging_power}
      \mathcal{P}(t) &= \frac{d \mathcal{W}[\rho_S(t)]}{d t}.
\end{align}
 A quantum battery has a positive charging power when charging, whereas its charging power is negative when discharging. A zero charging power indicates that the battery is neither charging nor discharging. Further, for a single-qubit system, it has been shown that an interesting relationship exists between charging and instantaneous power, with both connected by a non-zero factor~\cite{yadav2025thermo}.
\subsubsection{Average (dis-)charging power}
 The concept of average (dis-)charging power provides a useful measure of a quantum battery's performance over a given time interval. It is defined as the change in ergotropy during the period when the battery is being (dis-)charged, divided by the corresponding time duration during which the instantaneous charging power is negative (positive). Accordingly, the average (dis-)charging power over the interval $[t_i, t_f]$ can be expressed as
\begin{align}\label{eq_avg_charging_power}
\overline{\mathcal{P}} = \frac{\mathcal{W}(t_f) - \mathcal{W}(t_i)}{t_f - t_i}.
\end{align}
Here, $\mathcal{W}(t_f)$ and $\mathcal{W}(t_i)$ denote the ergotropy at the final and initial times, respectively.
This quantity indicates the net capacity with which the battery gains or loses its ergotropy, on average, over a given time interval.

\subsection{Description of the models envisaged as a quantum battery}\label{sec-models}
Here, we describe the models that specify the charger and the quantum battery, and categorize them based on different types of interaction.

\subsubsection{A model of two central spins immersed in spin baths}\label{sec_collective_central-spin-model}

We consider a model consisting of two coupled central spins (qubits), each interacting with its own local thermal spin bath, as in~\cite{PhysRevA.106.032435, devvrat_central_spin_2}. The two baths are independent, that is, they do not interact with each other, and each bath is an ensemble of identical spins. The total system Hamiltonian $H_{CS}$ is given by
\begin{align}
H_{CS} &= H_{S_1} + H_{S_2} + H_{S_1 S_2} + H_{B_1} + H_{B_2} + H_{S_1 B_1} + H_{S_2 B_2}
\end{align}
where
$H_{S_l}= \frac{\hbar\omega_1}{2}\sigma^0_{lz}$ ($l = 1, 2$),  
$H_{S_1S_2} = \hbar V_{S_1S_2}$,
$H_{B_1} = \frac{\hbar\omega_a}{2M} \sum^M_{i=1} \sigma^i_{1z}$, 
$H_{B_2} = \frac{\hbar\omega_b}{2N} \sum^N_{i=1}  \sigma^i_{2z}$, 
$H_{S_1B_1}= \frac{\hbar\epsilon_1}{2\sqrt{M}} \sum^M_{i=1}(\sigma^0_{1x}\sigma^i_{1x} + \sigma^0_{1y}\sigma^i_{1y})$, and 
$H_{S_2B_2} = \frac{\hbar\epsilon_2}{2\sqrt{N}}\sum^N_{j=1}(\sigma^0_{2x}\sigma^j_{2x} + \sigma^0_{2y}\sigma^j_{2y})$. 
Here, \(\sigma^i_{lk}\) or \(\sigma^j_{lk}\) \((k =x,y,z; l=1,2)\) are the standard Pauli spin matrices corresponding to $i$-th or $j$-th spin of the $l-$th bath and $\sigma^0_{lk}$ $(k = x,y,z; l=1,2)$ corresponds to the Pauli spin matrices for the $l-$th central spin. 
\(\omega_1\) and \(\omega_2\) are the transition frequencies of the two central spins. The interaction between the central spins is given by $V_{S_1S_2}$, specified below. Further, $\omega_a$ and $\omega_b$ are the bath frequencies of the two spin baths, and \(\epsilon_l\)'s are the uniform interaction strengths between the central spins and their corresponding spin baths. $M$ and $N$ are the number of spins in the two spin baths. 

Now, using the collective angular momentum operators $J_{lk} = \frac{1}{2}\sum_i\sigma_{lk}^i$ (with $k = x, y, z; l = 1, 2$),  the Hamiltonians describing the interaction between each central spin and its corresponding bath $H_{S_lB_l}$ can be rewritten as
\begin{align}
    H_{S_1B_1} &= \frac{\hbar \epsilon_1}{\sqrt{M}}\left(\sigma_{1x}^0J_{1x} + \sigma_{1y}^0J_{1y}\right),\nonumber \\
    H_{S_2B_2} &= \frac{\hbar \epsilon_2}{\sqrt{N}}\left(\sigma_{2x}^0J_{2x} + \sigma_{2y}^0J_{2y}\right),
\end{align}
and the bath Hamiltonians $H_{B_l}$ can be rewritten as
\begin{align}
    H_{B_1} = \hbar \omega_a \frac{J_{1z}}M, && \text{and} &&
    H_{B_2} = \hbar \omega_b \frac{J_{2z}}N .
\end{align}

Considering the joint initial state of the system and bath as $\rho_{SB}(0) = \rho_{S_1S_2}(0) \otimes \rho_{B_1}(0)\otimes\rho_{B_2}(0)$ (where thermal Gibbs state $\rho_{B_l}(0) = e^{-\beta_l H_{B_l}}/{\rm Tr}[e^{-\beta_l H_{B_l}}]$ is taken as the initial state of each bath), we find the reduced dynamics of the two central spin system using the unitary evolution of the joint system bath 
\begin{align}\label{Eq_system_evolution_two_central_spin_battery}
    \rho_{S_1S_2} = {\rm Tr}_{B_1B_2}\left(e^{-iH_{CS}t/\hbar}\rho_{SB}(0)e^{iH_{CS}t/\hbar}\right).
\end{align}

We shall analyze this model by considering two distinct types of interactions $V_{S_1S_2}$ between the central spins. The first is the Heisenberg XXX interaction~\cite{Heisenberg1928, Bethe1931, Mattis1981, spin_chain_primer_nepomechie} $V_{S_1S_2}^{\rm XXX} = g_{12}\left(\sigma^0_{1x} \otimes \sigma^0_{2x} + \sigma^0_{1y} \otimes \sigma^0_{2y} + \sigma^0_{1z} \otimes \sigma^0_{2z}\right)$. The second is the antisymmetric Dzyaloshinskii–Moriya (DM) interaction~\cite{Moriya, DZYALOSHINSKY1958}, given by $V_{S_1S_2}^{\rm DM} = g_{12,z}\left(\sigma^0_{1x} \otimes \sigma^0_{2y} - \sigma^0_{1y} \otimes \sigma^0_{2x}\right) + g_{12, x}\left(\sigma^0_{1y} \otimes \sigma^0_{2z} - \sigma^0_{1z} \otimes \sigma^0_{2y}\right) + g_{12, y}\left(\sigma^0_{1z} \otimes \sigma^0_{2x} - \sigma^0_{1x} \otimes \sigma^0_{2z}\right)$, where $g_{12, x} = g_{12, y} = g_{12,z} = g_{12}$ denotes the strength of the interaction. The two types of interaction are chosen to observe the effect of a symmetric and anti-symmetric exchange between the qubits of the batteries.

\begin{figure}
    \centering
    \includegraphics[width=1\linewidth]{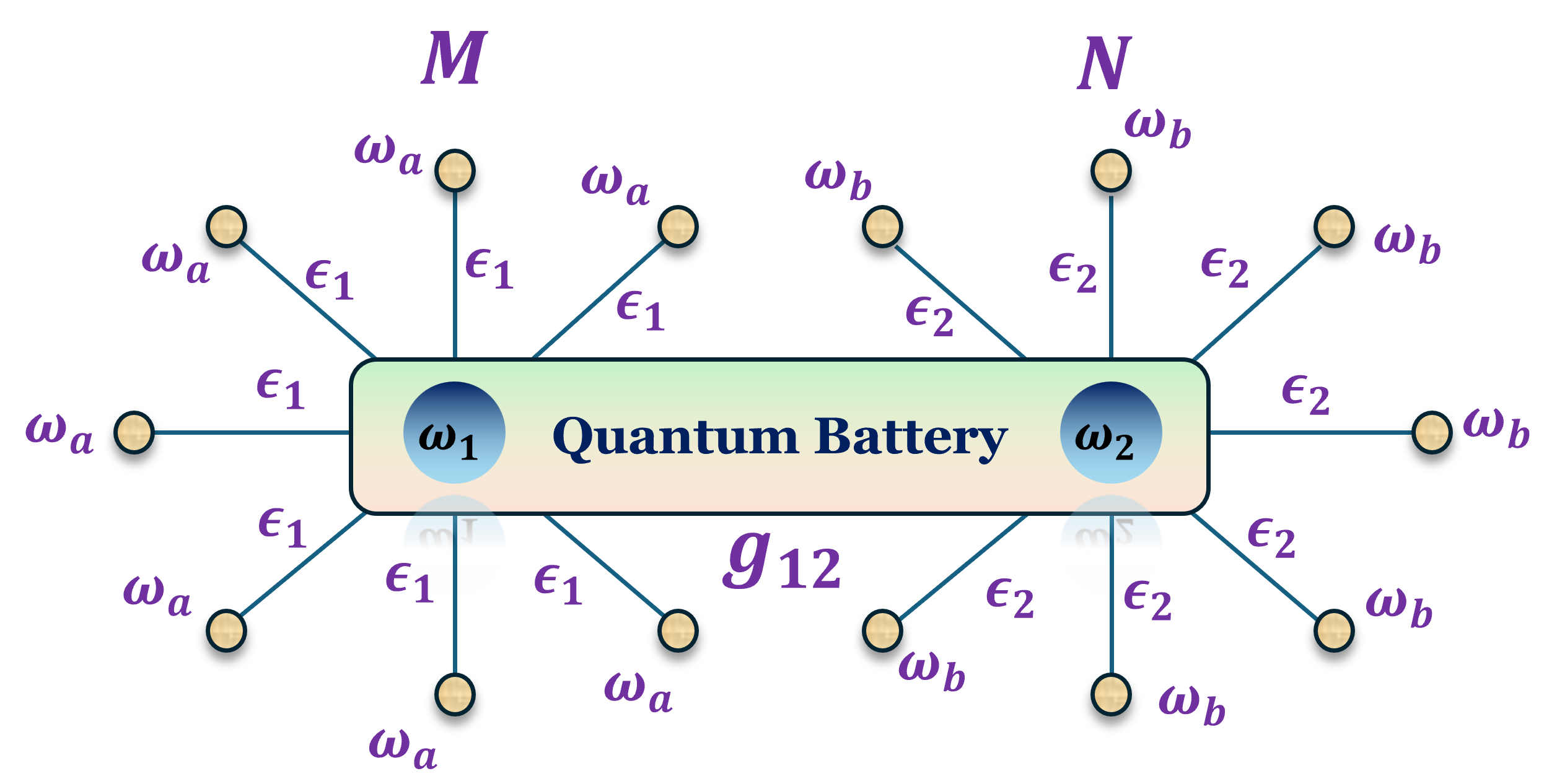}
    \caption{A schematic diagram of the two-qubit collective central spin quantum battery.}
    \label{fig_two_central_spin_collective_quantum_battery}
\end{figure}

In this model, the two central spins are collectively envisioned as a quantum battery, see Fig~\ref{fig_two_central_spin_collective_quantum_battery}, and their corresponding spin baths act as a dissipator or charger. Due to this, we call this setup a two-qubit collective central spin battery.
The impact of both the XXX and DM interactions on the performance of this quantum battery is analyzed in Sec.~\ref{sec_dynamics_two_qubit_collective_cs_battery}.  
 
\subsubsection{Two qubit collective decoherence model} \label{sec_two-qubit-decoherence-model}

Here, a dissipative interaction model of two qubits (two-level atomic system) interacting with the squeezed bath~\cite{BANERJEE2010} via dipole interaction~\cite{FICEK2002} is considered. The Hamiltonian of the total system is given by~\cite{FICEK2002, BANERJEE2010}
\begin{align}
\begin{split}
H_{\rm TQCD} &= H_S + H_B + H_{SB} \\
&= \frac{1}{2}\sum^2_{l=1} \hbar \omega_l \sigma^z_l + \sum_{\vec{k}s} \hbar\omega_k 
\left(b^{\dag}_{\vec{k}s} b_{\vec{k}s} + \frac{1}{2}\right) \\ &- i\hbar 
\sum_{\vec{k}s} \sum^2_{l=1} \left[\vec{\mu}_l \cdot 
\vec{g}_{\vec{k}s}(\vec{r}_l)\left(\sigma^{+}_{l} + 
\sigma^{-}_{l}\right)b_{\vec{k}s} - h.c.\right],
\end{split}
\end{align}
where $H_S$, $H_B$, and $H_{SB}$ are the system, bath, and system-bath interaction Hamiltonians, respectively, and \(\sigma^{+}_{l} = \frac{1}{2}\left(\sigma^x_l + i\sigma^y_l\right) = \ket{e_l}\bra{g_l}\) and \(\sigma^{-}_{l} = \frac{1}{2}\left(\sigma^x_l - i\sigma^y_l\right) = \ket{g_l}\bra{e_l}\) are the standard atomic raising and lowering operators corresponding to qubit $l$. $\omega_l$ is transition frequency for qubit $l$. The transition dipole moments \(\vec{\mu}_l\) depend on the atomic positions \(\vec{r}_l\). The operators \(b^{\dagger}_{\vec{k}s}\) and \(b_{\vec{k}s}\) are the bosonic
creation and annihilation operators, respectively, for the bath (field) mode $\vec{k}s$ with the wave vector \(\vec{k}\), polarization index \((s =1,2)\), and frequency \(\omega_k\). The system-bath (S-B) coupling constant is given by~\cite{FICEK2002, BANERJEE2010}
\begin{align*}
\vec{g}_{\vec{k}s}(\vec{r}_l) = \bigg(\frac{\omega_k}
{2\epsilon\hbar V} \bigg)^{\frac{1}{2}} \vec{e}_{\vec{k}s} 
e^{i\vec{k} \cdot \vec{r}_l} ,
\end{align*}
where \(V\) is the normalization volume and \(\vec{e}_{\vec{k}s}\) is the unit polarization vector of the field. 
The density matrix describing the reduced dynamics of the system, using Born-Markov and rotating wave approximations, is given by the master equation of the form ~\cite{BANERJEE2010, FICEK2002}
\begin{align}  \label{master-eqn}
\frac{d\rho}{dt} &= -\frac{i}{\hbar}[\tilde H_S,\rho] \nonumber \\
&- \frac{1}{2}\sum^{2}_{i,j=1} 
\Gamma_{ij}\left[1 + \tilde{N}\right]\left(\rho \sigma^{+}_{i} \sigma^{-}_{j} + \sigma^{+}_{i} \sigma^{-}_{j}\rho -2\sigma^{-}_{j}\rho\sigma^{+}_{i}\right) \nonumber \\
&- \frac{1}{2}\sum^{2}_{i,j=1} \Gamma_{ij} \tilde{N}\left(\rho \sigma^{-}_{i} \sigma^{+}_{j} + \sigma^{-}_{i} \sigma^{+}_{j}\rho -2\sigma^{+}_{j}\rho \sigma^{-}_{i}\right) \nonumber \\
&+ \frac{1}{2}\sum^{2}_{i,j=1} \Gamma_{ij} \tilde{M}\left(\rho \sigma^{+}_{i} \sigma^{+}_{j} + \sigma^{+}_{i} \sigma^{+}_{j}\rho -2\sigma^{+}_{j}\rho \sigma^{+}_{i}\right) \nonumber \\
&+ \frac{1}{2}\sum^{2}_{i,j=1} \Gamma_{ij} \tilde{M}^{*}\left(\rho \sigma^{-}_{i} \sigma^{-}_{j} + \sigma^{-}_{i}\sigma^{-}_{j}\rho -2\sigma^{-}_{j}\rho \sigma^{-}_{i}\right),
\end{align}
where
\begin{align}
\tilde{N} &= N_{th}(\cosh^{2} (r) + \sinh^{2} (r)) + \sinh^{2} (r),~~~~\text{and}  \\
\tilde{M} &= -\frac{1}{2}\sinh{(2r)}e^{i\Phi} (2N_{th} +1) \equiv Re^{i\Phi(\omega_0)},
\end{align}
with \(\omega_0 = \frac{\omega_1 + \omega_2}{2}\). \(N_{th} = \frac{1}{e^{\frac{\hbar\omega}{K_{B}T}} -1 }\) describes the Planck distribution, providing the number of thermal photons at the frequency \(\omega\) and \(r,\Phi\) are the squeezing parameters. The squeezed reservoir is assumed to be broadband, with squeezing bandwidths much larger than the atomic linewidths, such that the squeezing can be treated as frequency independent in the vicinity of the system transition frequency $\omega_0$. The carrier frequency of the squeezed field is taken to be resonant with the atomic transition. The explicit form of the system Hamiltonian used in Eq.~\eqref{master-eqn} is given by~\cite{FICEK2002, BANERJEE2010}
\begin{align}
\tilde{H}_{S} = \frac{\hbar}2\sum^{2}_{l=1}\omega_l \sigma^{z}_l + \hbar\sum^{2}_{\substack{i,j \\ (i\neq 
j)}}\Omega_{ij} \sigma^{+}_{i} \sigma^{-}_{j},
\end{align}
where 
\begin{align}
\Omega_{ij} &= \frac{3}{4}\sqrt{\Gamma_i \Gamma_j}\left[-\left\{1 - (\hat{\mu}\cdot \hat{r}_{ij})^2\right\} 
\frac{\cos{(k_0 r_{ij})}}{k_0 r_{ij}}\right. \nonumber \\ 
&\left.+ \left\{1 - 3(\hat{\mu}\cdot 
\hat{r}_{ij})^2\right\}\left\{\frac{\sin{(k_0 r_{ij})}}{(k_0 r_{ij})^2} + \frac{\cos{(k_0 r_{ij})}}
{(k_0 r_{ij})^3}\right\}\right],
\end{align}
with \(\hat{\mu} = \hat{\mu}_1 = \hat{\mu}_2\) being the unit vectors along the atomic transition dipole moments and $\hat{r}_{ij}$ is the unit vector along \(\vec{r}_{ij} = \vec{r}_i - \vec{r}_j\). Also, \(k_0 = \omega_0 /c \) and 
\(r_{ij} = |\vec{r}_{ij}|\). 
The wave vector is given by \(k_0 = \frac{2\pi}{\lambda_0}\), where \(\lambda_0\) is the resonant wavelength. The term $k_0r_{ij} \sim \frac{r_{ij}}{\lambda_0}$ in the above equation denotes a ratio between the interqubit distance and the resonant wavelength. This term classifies the system's dynamics into two distinct regimes: independent and collective decoherence. In the independent decoherence regime, where $k_0r_{ij}\ge1$, each qubit experiences the environment independently. Conversely, as $k_0r_{ij}\to 0$, the qubits are sufficiently close to each other, experiencing the environment collectively, thus referred to as the collective decoherence regime. Essentially, in the collective decoherence regime, the bath's correlation length, determined by $\lambda_0$, is longer than the distance between the qubits, $r_{ij}$.
Further, in the above equation, the spontaneous emission rate \(\Gamma_i\) is given by
\begin{align}
 \Gamma_i = \frac{\omega_{i}^{3}\mu^{2}_{i}}{3\pi\epsilon\hbar c^3},
\end{align}
while ~\(\Gamma_{ij} = \Gamma_{ji} = \sqrt{\Gamma_i \Gamma_j }F(k_0 r_{ij})\), where \(i\neq j\)
with
\begin{align*}
F(k_0 r_{ij}) &=  \frac{3}{2}\left[\left\{1 - (\hat{\mu}\cdot \hat{r}_{ij})^2\right\} 
\frac{\sin{(k_0 r_{ij})}}{k_0 r_{ij}}\right. \nonumber \\ 
&\left.+ \left\{1 - 3(\hat{\mu}\cdot 
\hat{r}_{ij})^2\right\}\left\{\frac{\cos{(k_0 r_{ij})}}{(k_0 r_{ij})^2} - \frac{\sin{(k_0 r_{ij})}}
{(k_0 r_{ij})^3}\right\}\right].
\end{align*}
\(\Gamma_{ij}\)'s represent collective spontaneous emission rates arising from the dissipative interaction of the multi-qubit system with the environment.

The two qubits are collectively modeled as a quantum battery, with the environment serving as a dissipator. To this effect, we coin the term for this battery as the collective decoherence battery.
The impact of the inter-atomic distance and the temperature of the bath on the quantum battery using ergotropy and their (in-)coherent parts is investigated in Sec.~\ref{sec_dynamics_two_qubit_collective_decoherence_battery}.

\subsubsection{Two central spin model in a charger-battery setup}\label{sec_central_spin_charger_battery_model}
\begin{figure}
    \centering
    \includegraphics[width=1\linewidth]{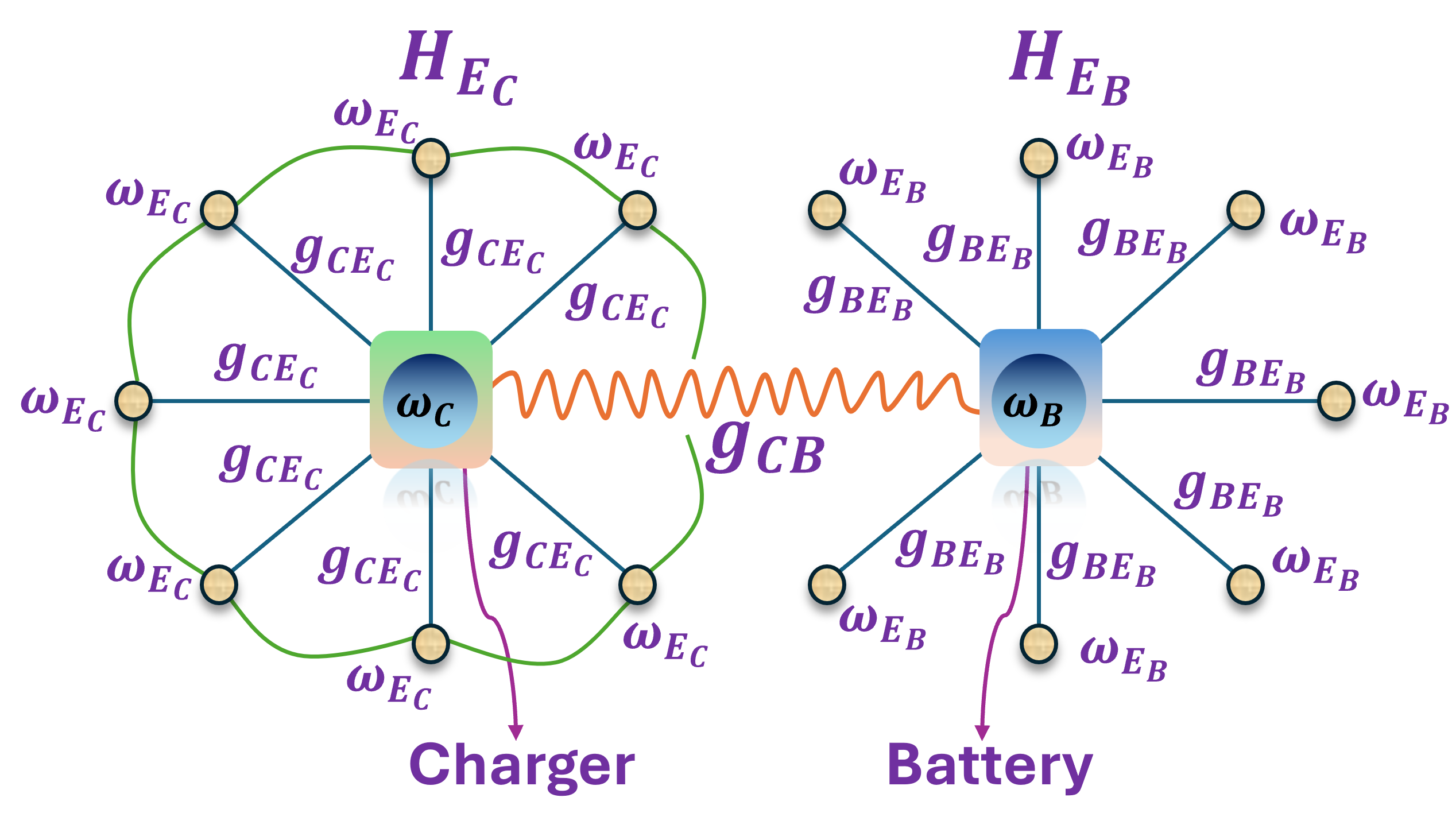}
    \caption{A schematic diagram of two central spins in a charger-battery setup.}
    \label{fig_two-qubit_central_spin_charger_battery_setup}
\end{figure}

The system consists of two central spins coupled to different spin baths. In this model, a central spin is modeled as a charger $H_C$ and is coupled to a spin-chain bath $H_{E_C}$, consisting of $N$ spins with nearest-neighbor interactions and an external magnetic field. The other central spin acts as a battery $H_B$ surrounded by a non-interacting spin bath $H_{E_B}$, consisting of $M$ spins, see Fig.~\ref{fig_two-qubit_central_spin_charger_battery_setup}.
The total Hamiltonian for this model is given by 
\begin{equation}
    H_{\rm C,B,E_C, E_B} = H_{C} + H_{B} + H_{CB} + H_{E_{C}} + H_{E_{B}} + H_{CE_{C}} + H_{BE_{B}},
\end{equation}
where (for $\hbar = 1)$
\begin{align}
    H_{C} &= \frac{\omega_{C}}{2}\sigma_{C}^z,~~~H_{B} = \frac{\omega_{B}}{2}\sigma_{B}^z,~~~H_{CB} = g_{CB} \left(\sigma_{C}^x \sigma_{B}^x + \sigma_{C}^y \sigma_{B}^y \right), \nonumber \\
    H_{E_{C}} &=  \frac{\omega_{E_{C}}}{2} \sum_{l = 1}^{N}\left[  \left(\frac{1 + \gamma}{2}\right)\sigma_{l}^{x} \sigma_{l + 1}^{x} + 
          \left( \frac{1 - \gamma}{2} \right)\sigma_{l}^{y}\sigma_{l + 1}^{y} - \lambda\sigma_{l}^{z} \right], \nonumber \\ 
    H_{E_{B}} &= \frac{\omega_{E_{B}}}{2} \sum_{k=1}^{M} \sigma_{k}^{z},~~~H_{CE_{C}}= g_{CE_{C}} \sum_{l=1}^{N} \left(\sigma_{C}^x \sigma_{l}^{x} + \sigma_{C}^y \sigma_{l}^{y}\right), \nonumber \\
    H_{BE_{B}} &= g_{BE_{B}} \sum_{k=1}^{M} \left(\sigma^x_B \sigma_{k}^{x} + \sigma^y_B \sigma_{k}^{y}\right) .
\end{align}
Here, $H_C$ and $H_B$ are the system Hamiltonians for the charger and the battery qubits, respectively, and $H_{E_{C}}$ is the bath Hamiltonian surrounding the charger and $H_{E_{B}}$ is the bath Hamiltonian surrounding the battery. Notably, the Hamiltonian for the bath surrounding the battery can be rewritten using the collective angular momentum operator $J_z = \frac{1}{2}\sum_k \sigma^z_k$ as $H_{E_B} = \omega_{E_B}J_z$. The environment interacting with the charger $H_{E_{C}}$ is considered to be an anisotropic XY spin-chain with the anisotropic parameter $\gamma$, and  $\lambda$ characterizes the strength of the transverse magnetic field applied in the $z$ direction. It should be noted that the last spin of the chain interacts with the first one, forming a closed spin chain. Furthermore, the interaction between the charger and battery is modeled by the Heisenberg XX interaction given by the Hamiltonian $H_{CB}$ with strength $g_{CB}$. The interaction Hamiltonian between the charger and its environments is given by $H_{CE_{C}}$ (with interacting strength $g_{CE_C}$), and that between the battery and its environment is given by $H_{BE_{B}}$ (with interaction strength $g_{BE_B}$). The transition frequencies for the battery and charger are given by \(\omega_{B}\) and $\omega_C$, respectively, and $\omega_{E_C}$ and $\omega_{E_B}$ are the transition frequencies of the bath spins in the baths surrounding the charger and the battery. 

The initial state of the composite system is taken to be $\rho(0) = \rho_{C}(0) \otimes \rho_{B}(0) \otimes \rho_{E_{C}}(0)  \otimes \rho_{E_{B}}(0)$. The initial state of the bath surrounding the charger (battery) $E_{C(B)}$ is taken to be the thermal state at temperature $T_{C(B)}$, that is, 
\begin{equation}
    \rho_{E_{C(B)}}(0) = \frac{\exp{\left(-H_{E_{C(B)}}/T_{C(B)}\right)}}{\Tr\left[ \exp{\left(-H_{E_{C(B)}}/{T_{C(B)}}\right)} \right]}.
\end{equation}
The reduced state of the charger-battery system is obtained by partially tracing their respective baths from the total dynamics as 
\begin{align}\label{rho_cbt}
    \rho_{CB}(t) = \Tr_{E_C, E_B}\left[U_{C, B, E_C, E_B}\left\{\rho(0)\right\}U_{C, B, E_C, E_B}^\dagger\right], 
\end{align}
where $U_{C, B, E_C, E_B} = \exp\left(-i H_{\rm C, B, E_C, E_B} t\right)$. Further, the reduced state of the battery at any time $t$ is given by $\rho_B(t) = \Tr_C\left[\rho_{CB}(t)\right]$. In principle, the initial states of both the charger and the battery can be taken arbitrarily. However, here, we consider the initial states of the charger and the battery as
\begin{align}\label{eq_charger_battery_initial_states}
    \rho_{C}(0) &= \begin{pmatrix}
        1 & 0 \\ 
        0 & 0
    \end{pmatrix}, \quad
    \rho_{B}(0) = \begin{pmatrix}
        1/2 & 1/2 \\ 
        1/2 & 1/2
    \end{pmatrix},
\end{align}
respectively. 

\section{Collective charging and discharging behavior}\label{sec_collective_battery}
Here, we consider the two-spin models discussed above and study the corresponding charging and discharging behavior. Both the spins in this scenario serve as a quantum battery, and their corresponding environment acts as a dissipating or recharging mechanism. 
\subsection{Two-qubit collective central spin battery}\label{sec_dynamics_two_qubit_collective_cs_battery}
\begin{figure}
    \centering
    \includegraphics[width=1\linewidth]{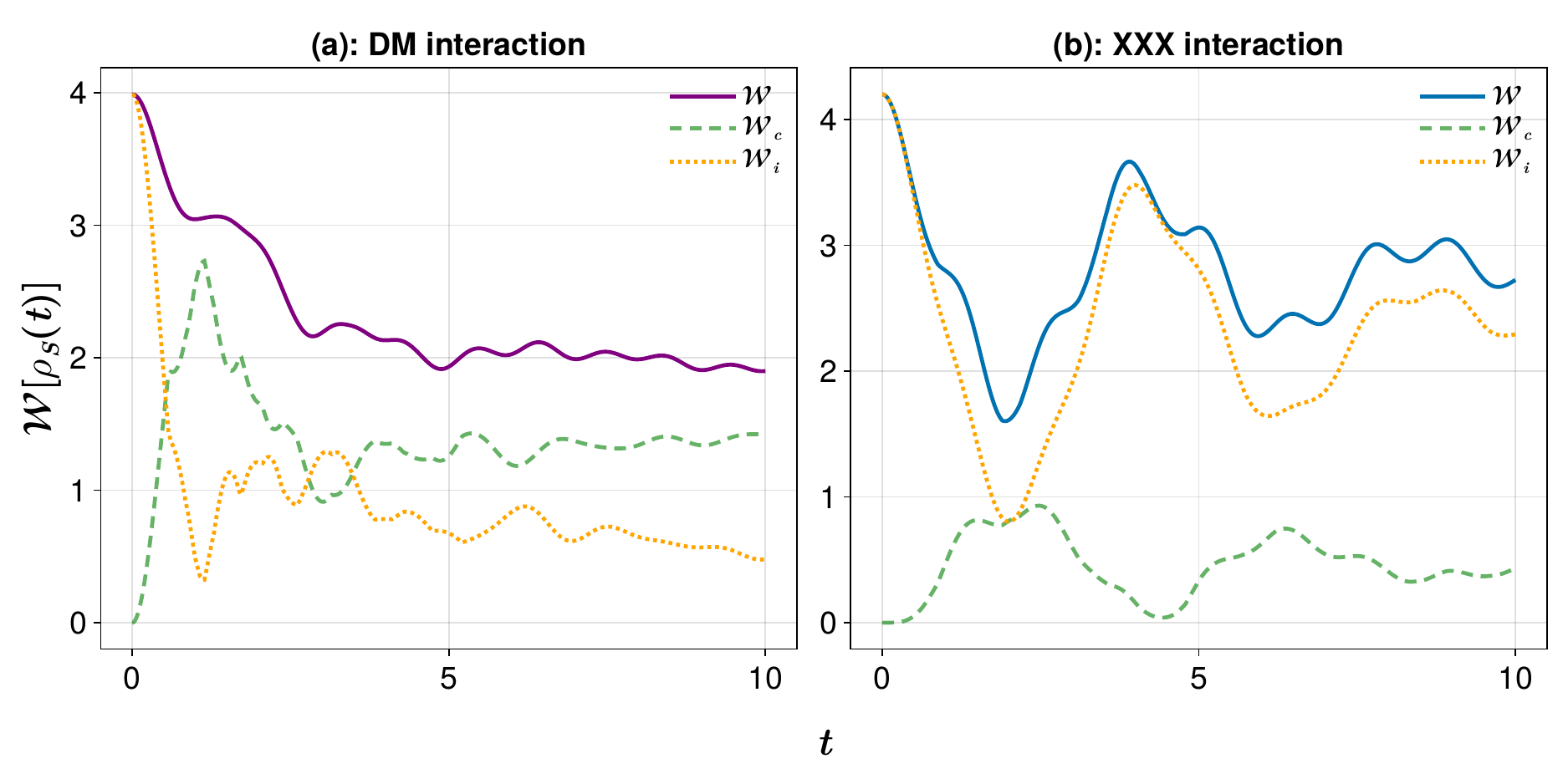}
    \caption{Variation of the ergotropy $\mathcal{W}[\rho_S(t)]$, coherent ergotropy $\mathcal{W}_c[\rho_S(t)]$, and incoherent ergotropy $\mathcal{W}_i[\rho_S(t)]$ with time (in natural units, where $\hbar = k_B = 1$) for the two central spin quantum battery for both (a) DM and (b) Heisenberg XXX inter-qubit interactions. The parameters are taken to be: $\omega_1= 1.15, \omega_2 = 1.25, \omega_a = 1.1, \omega_b = 1.2, g_{12} = 0.75, \epsilon_1 = \epsilon_2 = 0.5, \beta_a = 4, \beta_b = 1$, and $M= N= 8$.}
    \label{fig_ergotropy_two_central_spin_battery}
\end{figure}

\begin{figure}
    \centering
    \includegraphics[width=1\linewidth]{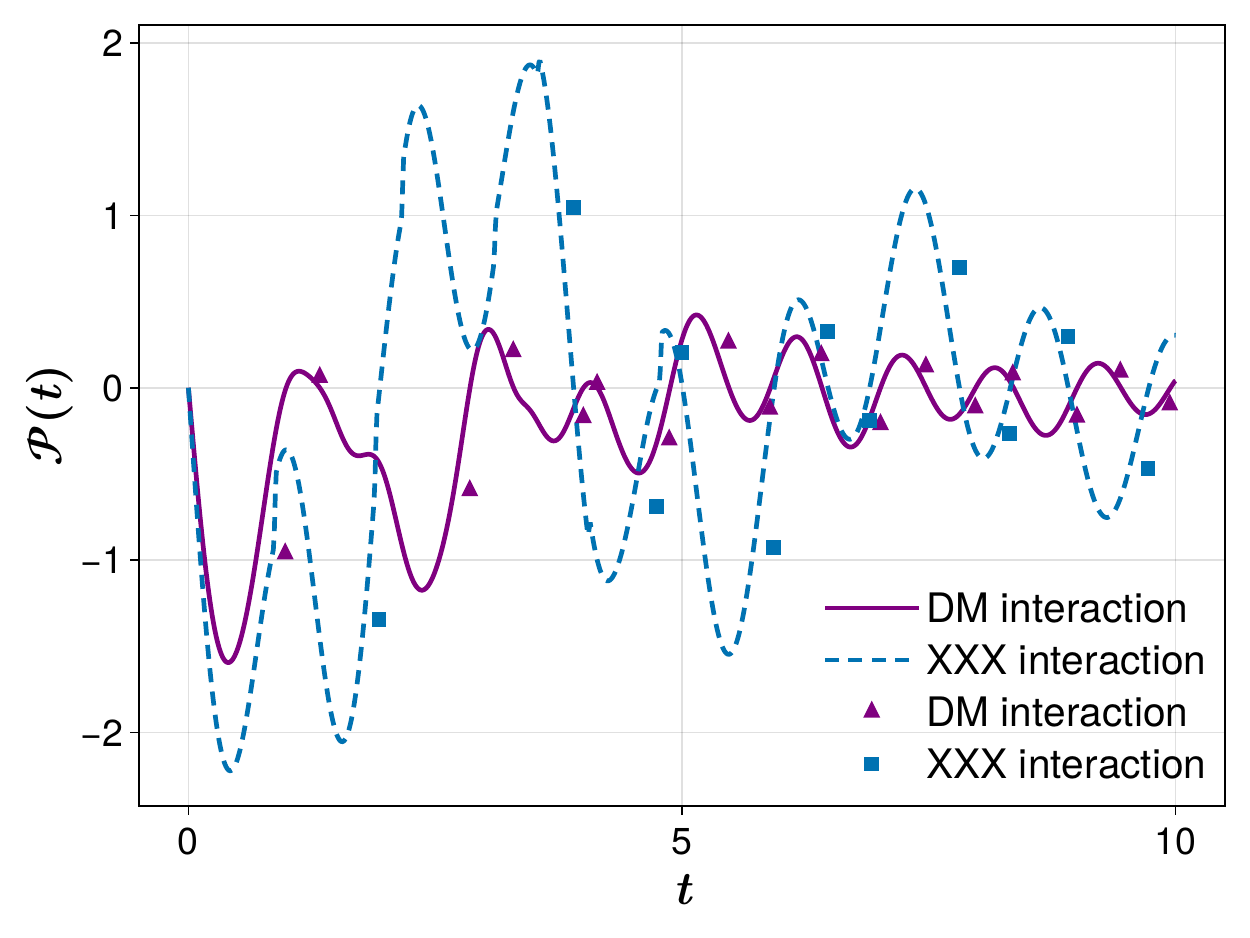}
    \caption{Variation of the charging power $\mathcal{P}(t)$, Eq.~\eqref{eq_charging_power}, and average (dis-)charging power $\overline{\mathcal{P}}$, Eq.~\eqref{eq_avg_charging_power}, with time (in natural units, where $\hbar = k_B = 1$) for the two central spin quantum battery for both DM and Heisenberg XXX interactions. The triangle and square markers show the average (dis-)charging power for DM and XXX inter-qubit interactions, respectively. The parameters are taken to be: $\omega_1= 1.15, \omega_2 = 1.25, \omega_a = 1.1, \omega_b = 1.2, g_{12} = 0.75, \epsilon_1 = \epsilon_2 = 0.5, \beta_a = 4, \beta_b = 1$, and $M= N= 8$.}
    \label{fig_power_two_central_spin_battery}
\end{figure}

We model the interaction between the two central spins using two different types of interactions, see Sec.~\ref{sec_collective_central-spin-model}. In the first case, the interaction between the two central spins is governed by the Heisenberg XXX interaction, and in the second case, the two central spins interact via the DM interaction, see below Eq.~\eqref{Eq_system_evolution_two_central_spin_battery}. Initially, the system is considered to be in the excited state $\ket{00}$, where, in the computational basis, $\ket{0} = \begin{pmatrix}
    1 \\ 0
\end{pmatrix}$ and $\ket{1} = \begin{pmatrix}
    0 \\ 1
\end{pmatrix}$ denote the excited and ground state of the system, respectively. Using the initial state of the system, we calculate the reduced state of the two-qubit central spin model as given by Eq.~\eqref{Eq_system_evolution_two_central_spin_battery}. The ergotropy, Eq.~\eqref{eq_ergotropy}, at each point of time, is calculated using this state of the system, and is depicted in Fig.~\ref{fig_ergotropy_two_central_spin_battery}.
It can be observed that the ergotropy decreases rapidly for the XXX interaction compared to the DM interaction, indicating a quicker initial discharge of the system for this interaction. However, the recharging, facilitated by the corresponding spin baths, helps regain the ergotropy for the XXX interaction, which takes higher values than the ergotropy for the DM interaction.
This highlights that even though the battery discharges quickly when XXX interaction is present, it recharges and maintains a higher amount of work that can be extracted from it. The coherent and incoherent parts of the ergotropy are further plotted for both DM and XXX interactions in Figs.~\ref{fig_ergotropy_two_central_spin_battery}(a) and ~\ref{fig_ergotropy_two_central_spin_battery}(b), respectively. In the case of the XXX interaction, the incoherent ergotropy contributes significantly to the ergotropy. In contrast, in the case of DM interaction, which is an antisymmetric, spin-orbit-type exchange~\cite{Albert_2023}, redistribution of populations toward lower-energy eigenstates occurs more rapidly, and hence the incoherent ergotropy remains lower. Later, the coherent ergotropy is the major contributor to the ergotropy in this case. This illustrates that the higher incoherent ergotropy is the reason behind a greater overall ergotropy for XXX inter-qubit interaction.
The corresponding charging power $\mathcal{P}(t)$, Eq.~\eqref{eq_charging_power}, and average (dis-)charging power $\overline{\mathcal{P}}$, Eq.~\eqref{eq_avg_charging_power}, for both types of inter-qubit interactions are plotted in Fig.~\ref{fig_power_two_central_spin_battery}. The positive charging power indicates charging of the quantum battery, and the discharging is indicated by the negative charging power. Furthermore, the average charging power is denoted by the corresponding markers with positive values, while the negative-valued markers depict the average discharging power.
The charging power for the XXX inter-qubit interaction initially takes higher negative values, indicating that it discharges more compared to the DM inter-qubit interactions. However, the charging power also takes higher positive values for XXX interaction, illustrating that it can deliver more work at these times than the battery with DM inter-qubit interactions. The battery with XXX interaction also has a higher average (dis-)charging power. This highlights that, on average, the battery with XXX interaction can deliver more power than the battery with DM interaction. 

\subsection{Two-qubit collective decoherence battery}\label{sec_dynamics_two_qubit_collective_decoherence_battery}
\begin{figure}
    \centering
    \includegraphics[width = 1\columnwidth]{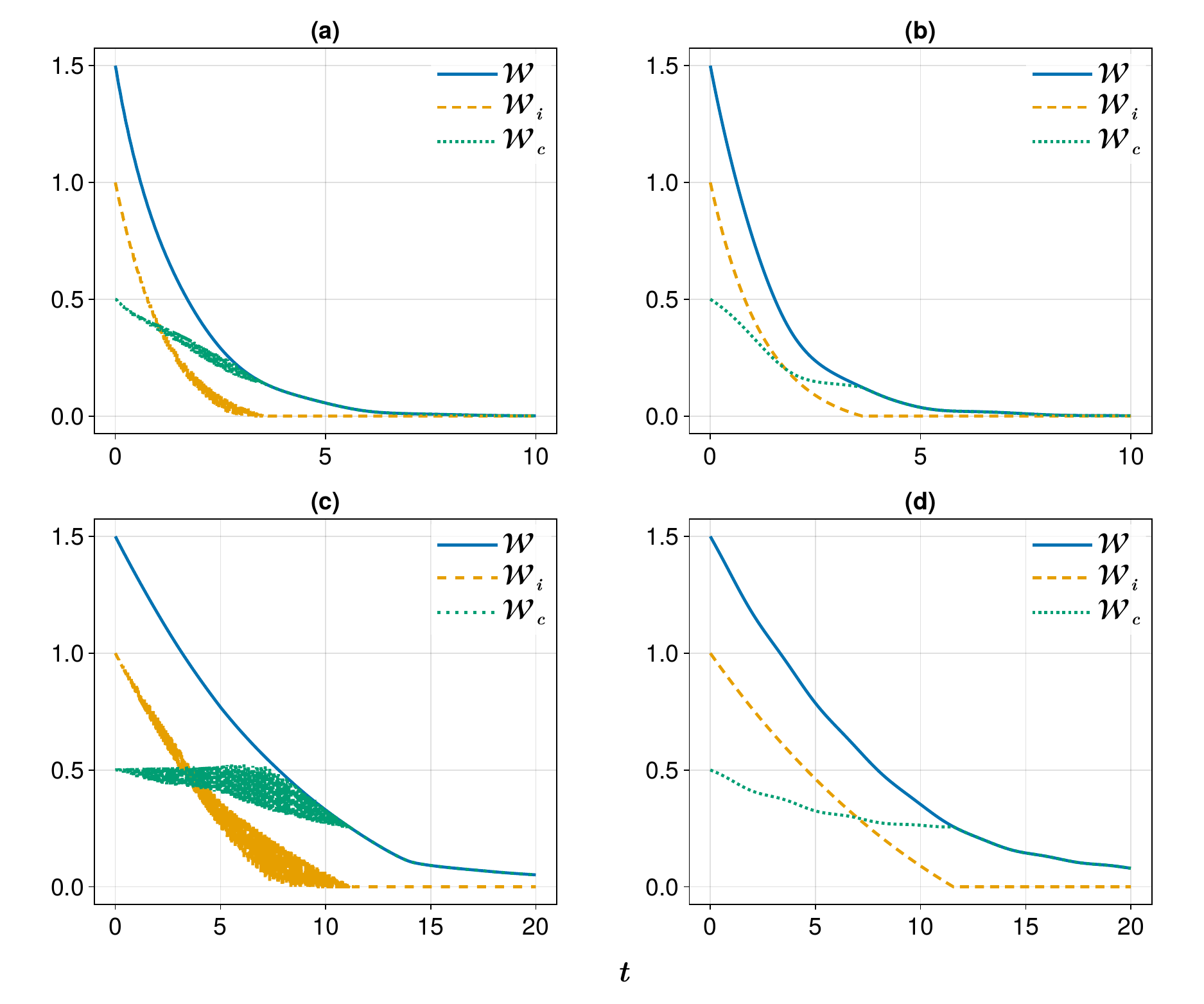}
   \caption{Ergotropy dynamics with time (in natural units, where $\hbar = k_B = 1$) for the two-qubit collective decoherence quantum battery. The initial two-qubit state is considered a product state $\ket{0+} = \frac{1}{\sqrt{2}}(\ket{00} + \ket{01})$. (a) and (b) correspond to a squeezed thermal bath at temperature $T=5$ with squeezing parameters $r = 0.5$ and $\Phi = \tfrac{\pi}{4}$, while Figures (c) and (d) show the vacuum bath case. Subplots (a) and (c) depict results for $k_{0}r_{ij}=0.1$ (collective decoherence), and (b) and (d) are for $k_{0}r_{ij}=1.2$ (independent dissipation). The other parameters are: $\omega_{1} = \omega_{2} = 1.0$, $\mu r_{ij} = 0$, $\Gamma_{1} = \Gamma_{2} = 0.05$.}
    \label{fig_Ergotropy_2_qubit_distance_model_time}
\end{figure}
Here, we analyze the performance of the two-qubit collective decoherence quantum battery, see Sec.~\ref{sec_two-qubit-decoherence-model}. In Fig.~\ref{fig_Ergotropy_2_qubit_distance_model_time}, we discuss the time evolution of ergotropy considering the effects of both the bath and interatomic distance. Both the coherent and incoherent components contribute to the total ergotropy, see Eq.~\eqref{eq_incoherent_ergotropy}. While the coherent part dominates over longer times, the incoherent contribution decays rapidly and becomes negligible. When the qubits are close enough to experience collective decoherence, revivals appear in both the coherent and incoherent ergotropy. These revivals are synchronized in such a way that the overall ergotropy still exhibits a monotonic decay, as shown in Figs.~\ref{fig_Ergotropy_2_qubit_distance_model_time}(a) and (c). In contrast, when the qubits undergo independent decoherence, both the (in-)coherent ergotropy decay monotonically, and so does the ergotropy. Further, a comparison between different bath conditions highlights the role of temperature. Under the influence of the squeezed thermal bath, the ergotropy decays faster than that in the vacuum bath case due to the finite effective temperature of the squeezed bath, which accelerates the dissipation of extractable work.

\begin{figure}
    \centering
    \includegraphics[width = 1\columnwidth]{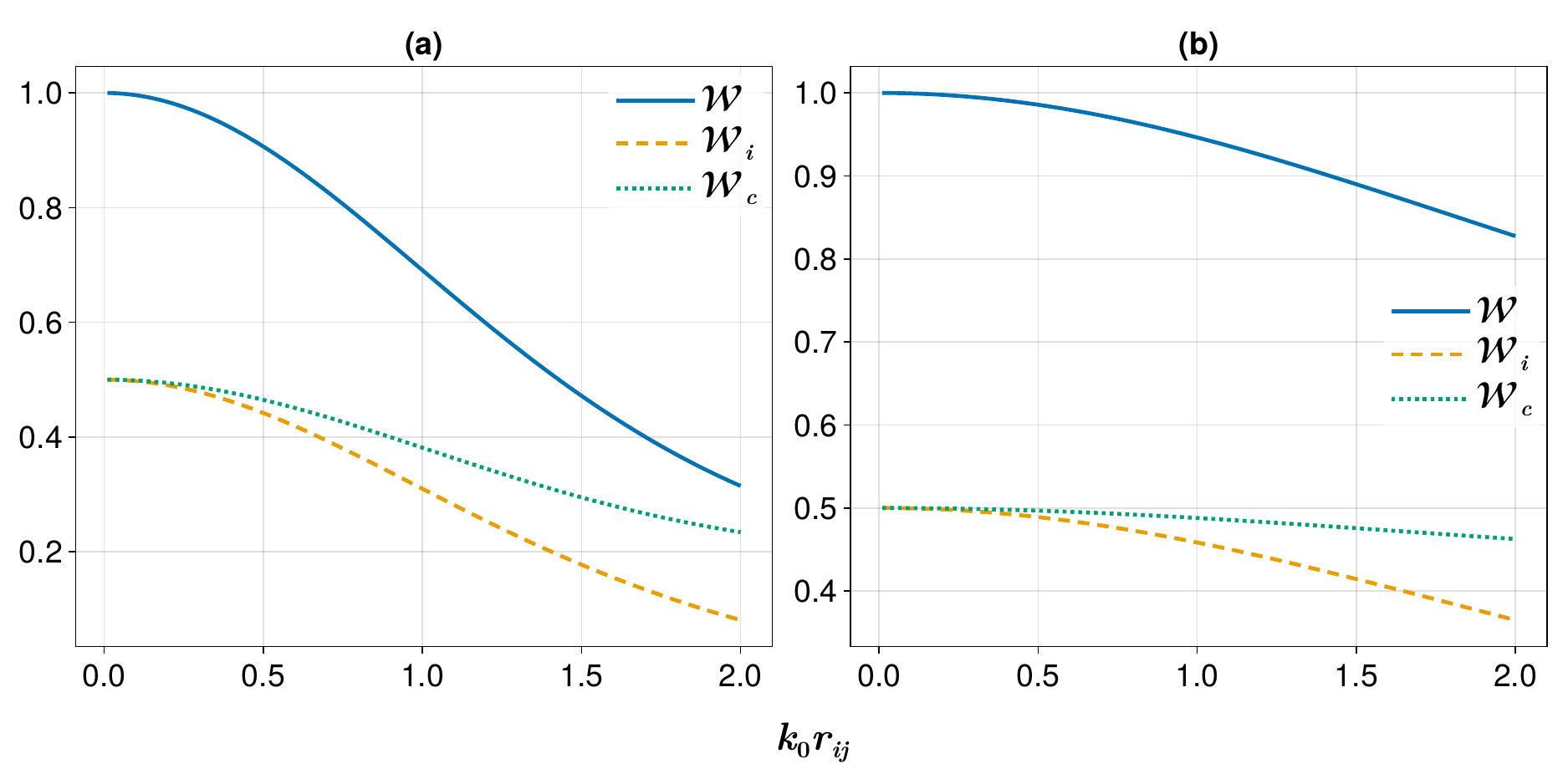}
   \caption{Variation of ergotropy with the interatomic distance $r_{ij}$ at a given time $t = 2$ for the two-qubit collective decoherence battery. The initial state of the two qubits is taken to be the Bell-state $\frac{1}{\sqrt{2}}\left(\ket{01} - \ket{10}\right)$. The squeezed thermal bath parameters are: (a) $T=5$, $r=0.5$, $\Phi = \frac{\pi}{4}$, and (b) $T=0.4$, $r=0.5$, $\Phi =\frac{\pi}{4}$. Further, $\omega_{1} = \omega_{2} = 1.0$, $\mu r_{ij} = 0$, $\Gamma_{1} = \Gamma_{2} = 0.05$. Natural units are taken, where $\hbar = k_B = 1$.}
    \label{fig_Ergotropy_2_qubit_distance_model_BS}
\end{figure}

Furthermore, we analyze the effect of interatomic distance on the evolution of ergotropy for qubits initially prepared in a Bell state. As shown in Figs.~\ref{fig_Ergotropy_2_qubit_distance_model_BS}(a) and (b), the total ergotropy, including both coherent and incoherent contributions, decreases very slowly under collective decoherence ($r_{ij} \leq 1$), particularly at low temperature, see Fig.~\ref{fig_Ergotropy_2_qubit_distance_model_BS}(b) where it remains nearly stationary. In contrast, for independent decoherence ($r_{ij} \geq 1$), the amount of extractable work from the battery decays faster as the incoherent part vanishes quickly, while the coherent part continues to contribute but also diminishes with time. Notably, in the low temperature case, Fig.~\ref{fig_Ergotropy_2_qubit_distance_model_BS}(b), the decay of ergotropy is slower compared to the high-temperature case in Fig.~\ref{fig_Ergotropy_2_qubit_distance_model_BS}(a).

\begin{figure}
    \centering
    \includegraphics[width = 1\columnwidth]{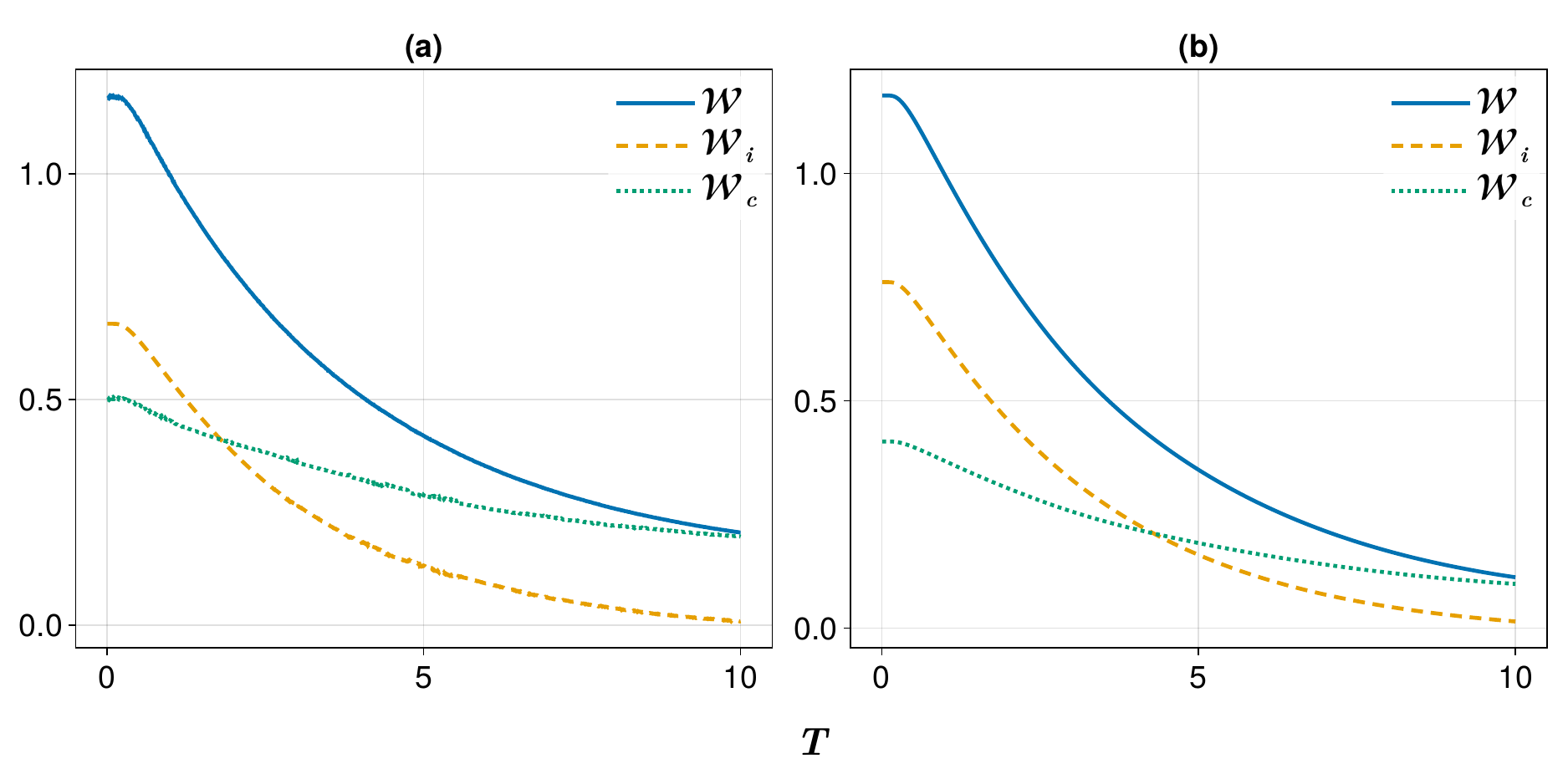}
   \caption{Variation of ergotropy with temperature for the 2-qubit system in interaction with a squeezed thermal bath at time $t=2$ with squeezing parameters $r = 0.5$ and $\Phi = \tfrac{\pi}{4}$. (a) depicts results for $k_{0}r_{ij}=0.05$ (collective decoherence) and (b) for $k_{0}r_{ij}=1.1$ (independent dissipation). The other parameters are: $\omega_{1} = \omega_{2} = 1.0$, $\mu r_{ij} = 0$, $\Gamma_{1} = \Gamma_{2} = 0.05$. Natural units are taken, where $\hbar = k_B = 1$.}
    \label{fig_Ergotropy_2_qubit_distance_model_Temp}
\end{figure}

In Fig.~\ref{fig_Ergotropy_2_qubit_distance_model_Temp}, we examine the dependence of ergotropy on temperature for both the collective and independent coherence, with the battery qubits initially prepared in the product state $\ket{0+}$. As shown in Figs.~\ref{fig_Ergotropy_2_qubit_distance_model_Temp}(a) and (b), increasing temperature enhances the dissipation of ergotropy into the environment, thereby reducing the amount of extractable work. For collective decoherence, the coherent contribution remains larger than in the case of independent decoherence, resulting in a comparatively slower decay of the total ergotropy. 

In Figs.~\ref{fig_Ergotropy_2_qubit_distance_model_time} and~\ref{fig_Ergotropy_2_qubit_distance_model_Temp}, the initial dominance of the incoherent ergotropy is due to the initial state of the system. Further, due to dissipation of the higher-energy population terms, the incoherent ergotropy drops rapidly. However, the coherence of the system remains, responsible for the crossover. At longer duration and higher temperatures, the coherent ergotropy becomes the overall ergotropy of the system.

\section{Two central spins as charger and battery}\label{sec_charger_battery_setup_investigation}
Here, the two central spins are considered in a charger-battery setup, with one modeled as a charging qubit and the other as a battery qubit, see Eq.~\eqref{eq_charger_battery_initial_states} and Sec.~\ref{sec_central_spin_charger_battery_model}. The charging central spin is coupled to an anisotropic spin-1/2 XY chain, whereas the battery is interacting with a spin bath of non-interacting qubits. For simplicity, we set $\gamma = 1$; in this case, the spin chain becomes the Ising model. For any value of  $\gamma$, quantum criticality occurs at the critical magnetic field strength $\lambda_{c} = 1$. It has been shown earlier that the quantum speed limit time (QSLT) has some strong imprint of the quantum phase transition for the XY model, even for a finite-sized environment, and exhibits noticeable anomalous behavior near the critical point, where the non-Markovianity (calculated using the Breuer-Laine-Piilo (BLP) measure~\cite{BLP_measure}) was also shown to be impacted by the phase transition~\cite{Wei2016}. In order to study the performance of the quantum battery near the critical point, we plot the ergotropy, energy, and power as a function of magnetic field $\lambda$ and time.

\begin{figure}
    \centering
    \includegraphics[width = 1\columnwidth]{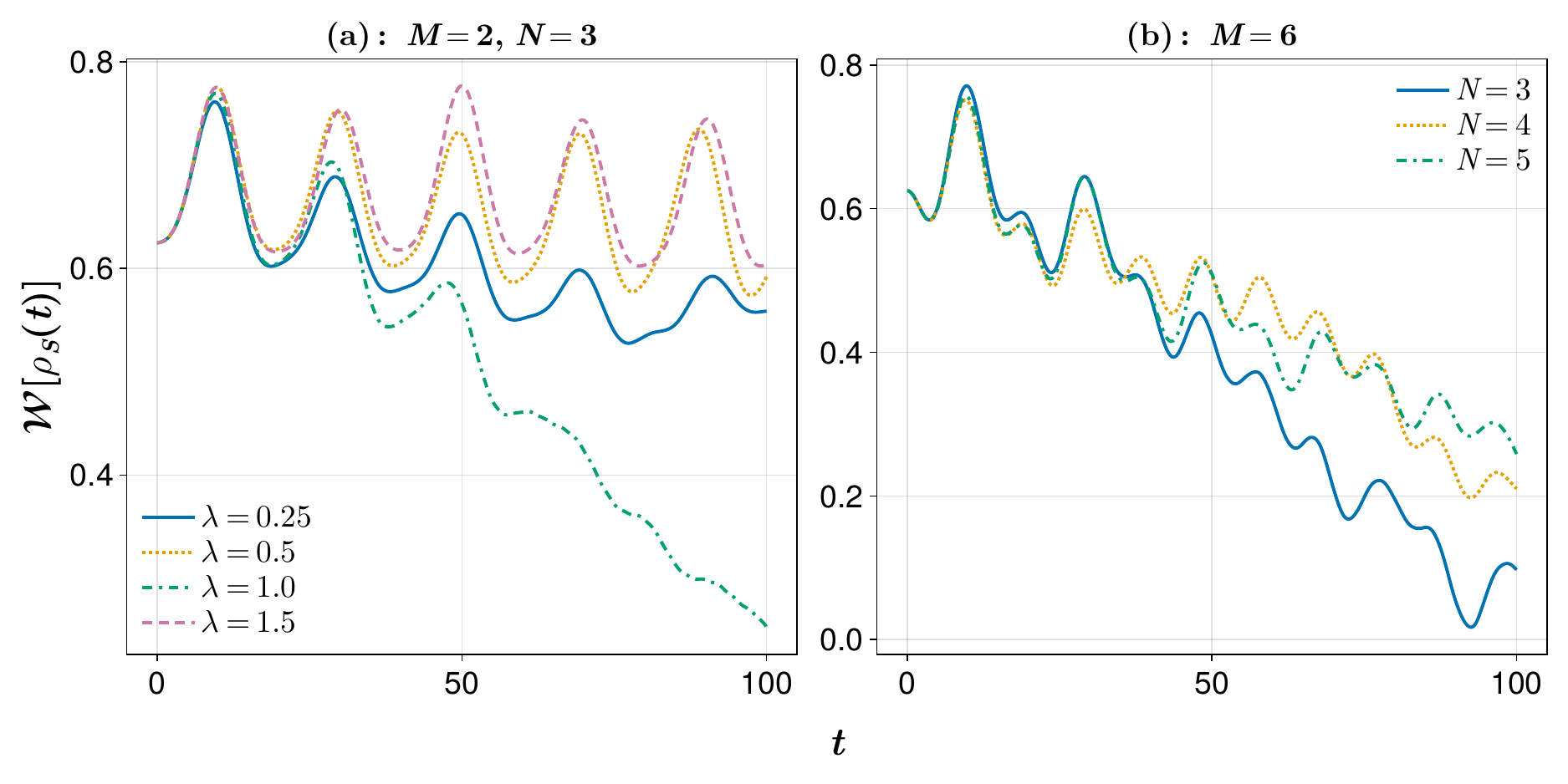}
    \caption{Variation of the ergotropy with time for the two central spin charger-battery model for different $\lambda$ in (a), and for different number of bath spins interacting with the charger in (b) at $\lambda = 1.0$. The parameters are: $\omega_{C} = 1.5$, $\omega_{B} = 1.25$, $g_{CB} = 0.05$, $g_{CE_{C}} = 0.04$, $g_{BE_{B}}=0.02$, $\omega_{E_{B}} = 0.6$, $\omega_{E_{C}} = 0.7$, $\gamma = 1$, $T_{C} = 0.5$, $T_{B} = 0.8$. In (a), $M=2$, $N=3$, and in (b) $M = 6$. Natural units are taken, where $\hbar = k_B = 1$.}
    \label{fig:Ergotropy_time_2CentralSpin_model_CB}
\end{figure}

\begin{figure}
    \centering
    \includegraphics[width = 1\columnwidth]{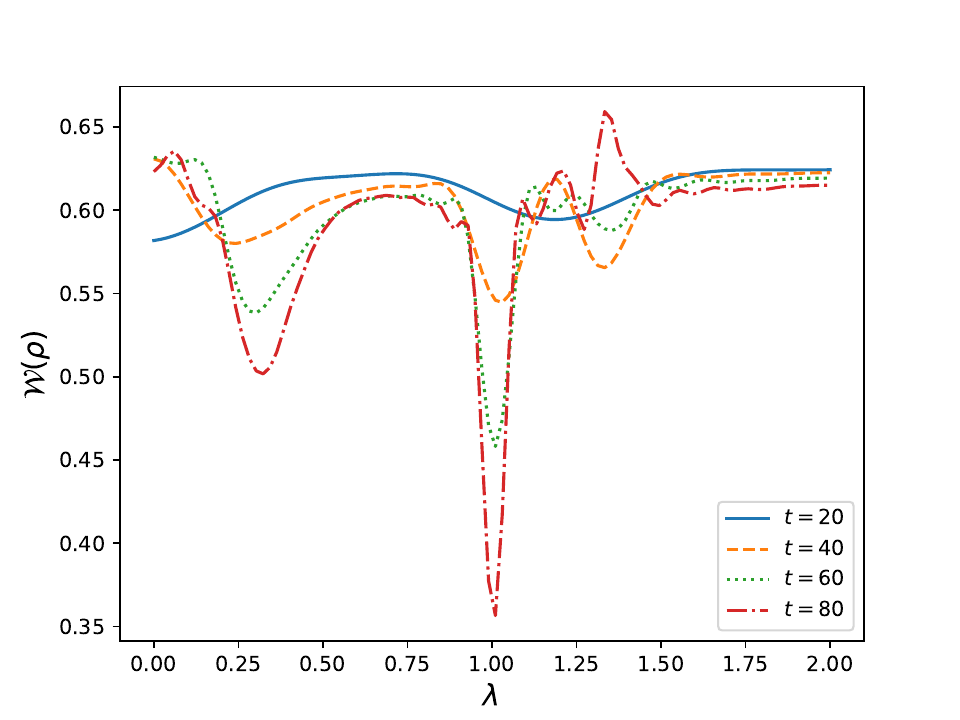}
    \caption{Variation of the ergotropy with $\lambda$ for the two central spin charger-battery model at different times. The parameters are: $\omega_{C} = 1.5$, $\omega_{B} = 1.25$, $g_{CB} = 0.05$, $g_{CE_{C}} = 0.04$, $g_{BE_{B}}=0.02$, $\omega_{E_{B}} = 0.6$, $\omega_{E_{C}} = 0.7$, $M=2$, $N=3$, $\gamma = 1$, $T_{C} = 0.5$, $T_{B} = 0.8$. Natural units are taken, where $\hbar = k_B = 1$.}
    \label{fig:Ergotropy_lamda_2CentralSpin_model_CB}
\end{figure}

\begin{figure}
    \centering
    \includegraphics[width = 1\columnwidth]{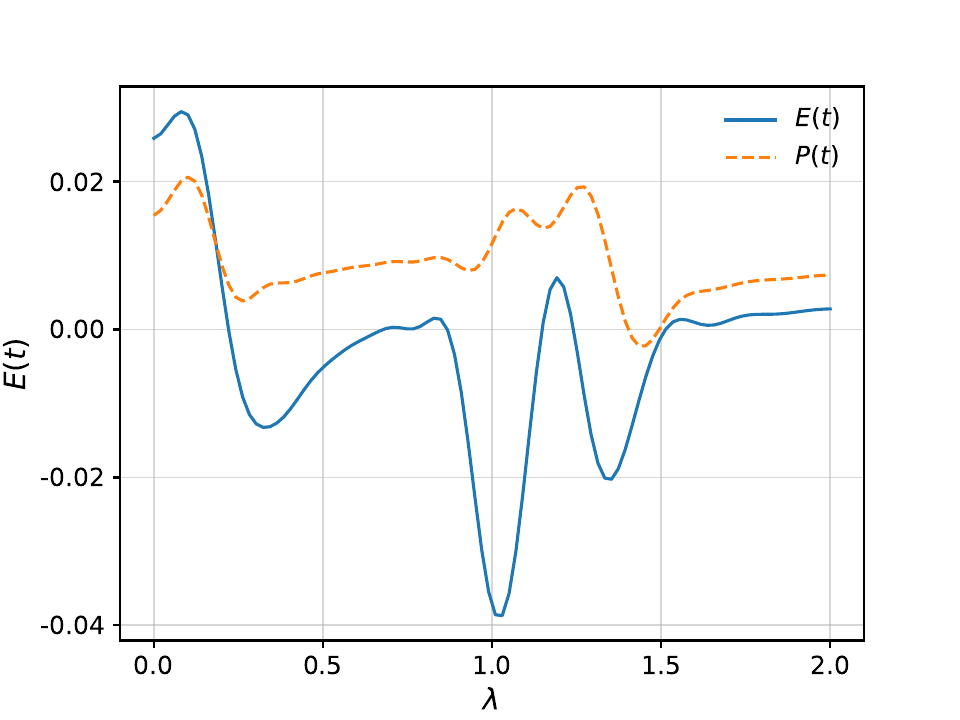}
    \caption{Variation of the energy and power of battery spin with magnetic field strength $\lambda$ for the two central spin charger-battery model at time $t = 40$. The parameters are: $\omega_{C} = 1.5$, $\omega_{B} = 1.25$, $g_{CB} = 0.05$, $g_{CE_{C}} = 0.04$, $g_{BE_{B}}=0.02$, $\omega_{E_{B}} = 0.6$, $\omega_{E_{C}} = 0.7$, $M=2$, $N=3$, $\gamma = 1$, $T_{C} = 0.5$, $T_{B} = 0.8$. Natural units are taken, where $\hbar = k_B = 1$.}
    \label{fig:Energyy_powerwith_lambda_2CentralSpin_model_CB}
\end{figure}

\begin{figure}
    \centering
    \includegraphics[width = 1\columnwidth]{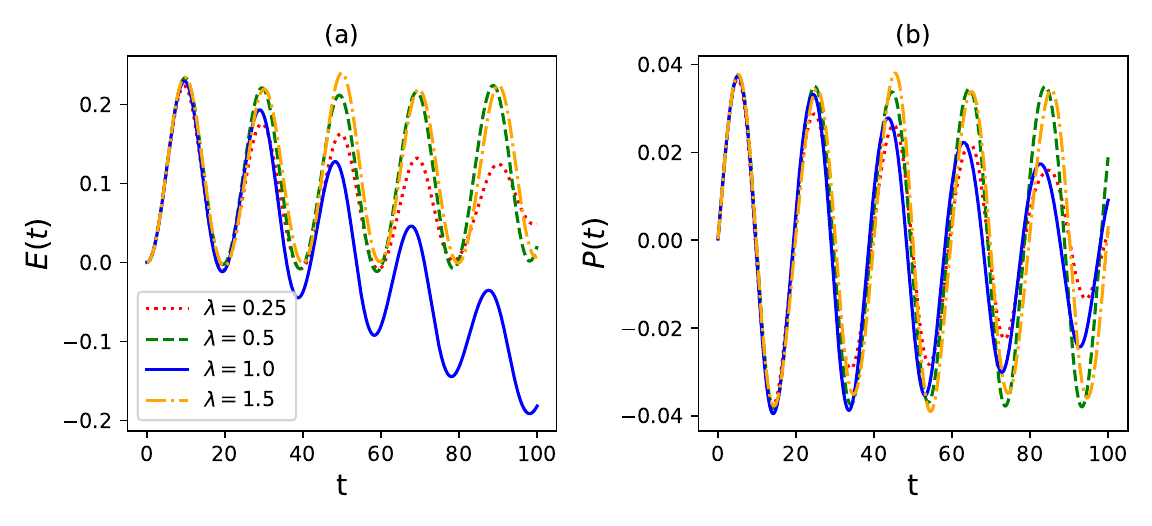}
    \caption{Variation of the energy $E(t)$ and power $P(t)$ of battery spin with time for the two central spin charger-battery model at different $\lambda$. The parameters are: $\omega_{C} = 1.5$, $\omega_{B} = 1.25$, $g_{CB} = 0.05$, $g_{CE_{C}} = 0.04$, $g_{BE_{B}}=0.02$, $\omega_{E_{B}} = 0.6$, $\omega_{E_{C}} = 0.7$, $M=2$, $N=3$, $\gamma = 1$, $T_{C} = 0.5$, $T_{B} = 0.8$. Natural units are taken, where $\hbar = k_B = 1$.}
    \label{fig:Energyy_powerwith_time_2CentralSpin_model_CB}
\end{figure}

\begin{figure}
    \centering
    \includegraphics[width = 1\columnwidth]{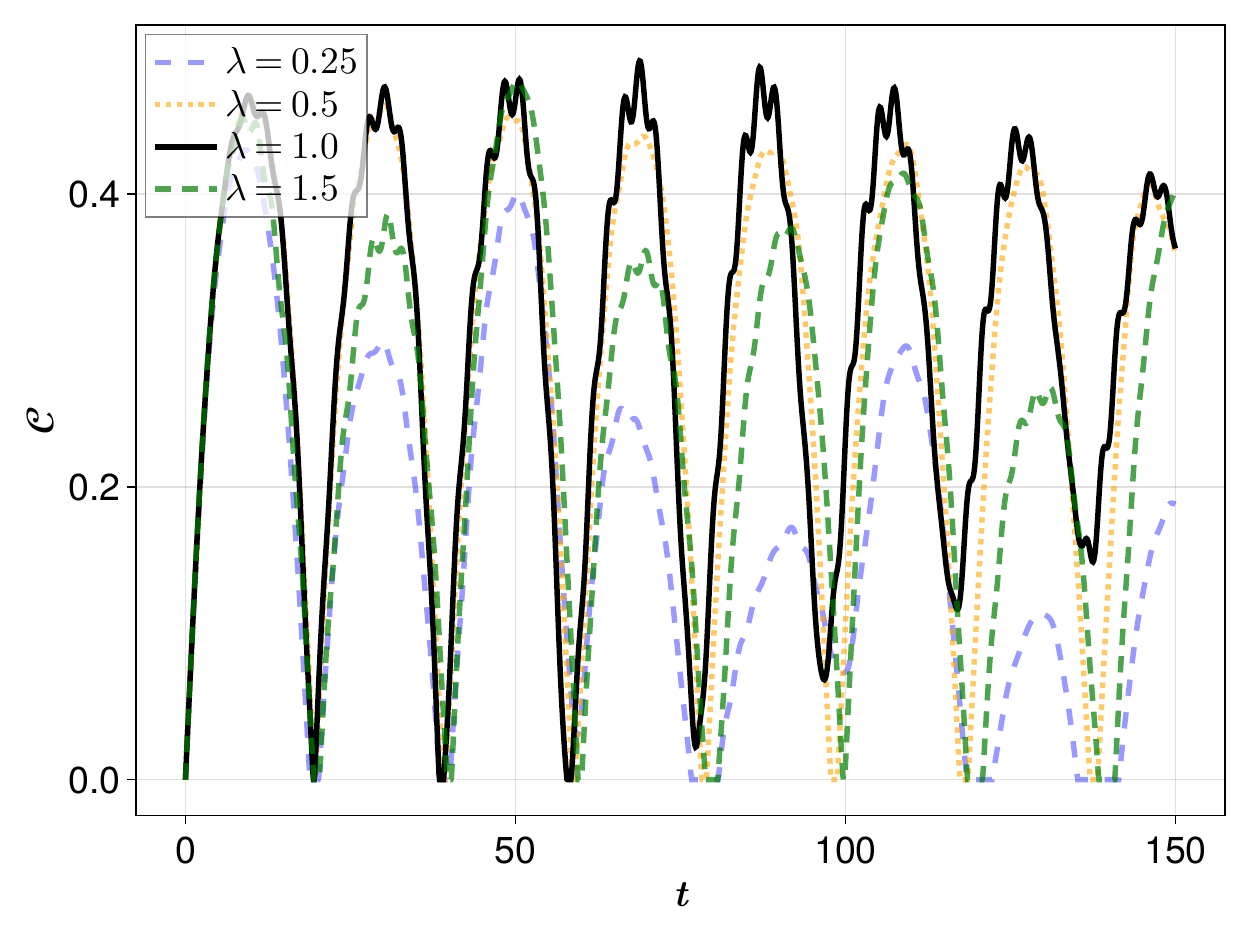}
    \caption{Variation of the concurrence $\mathcal{C}$ of charger-battery spin with time for the two central spin charger-battery model at different $\lambda$. The parameters are: $\omega_{C} = 1.5$, $\omega_{B} = 1.25$, $g_{CB} = 0.05$, $g_{CE_{C}} = 0.04$, $g_{BE_{B}}=0.02$, $\omega_{E_{B}} = 0.6$, $\omega_{E_{C}} = 0.7$, $M=2$, $N=3$, $\gamma = 1$, $T_{C} = 0.5$, $T_{B} = 0.8$. Natural units are taken, where $\hbar = k_B = 1$.}
    \label{fig:Concurrencewith_time_2CentralSpin_model_CB1}
\end{figure}

\begin{figure*}
    \centering
    \includegraphics[width=0.95\linewidth]{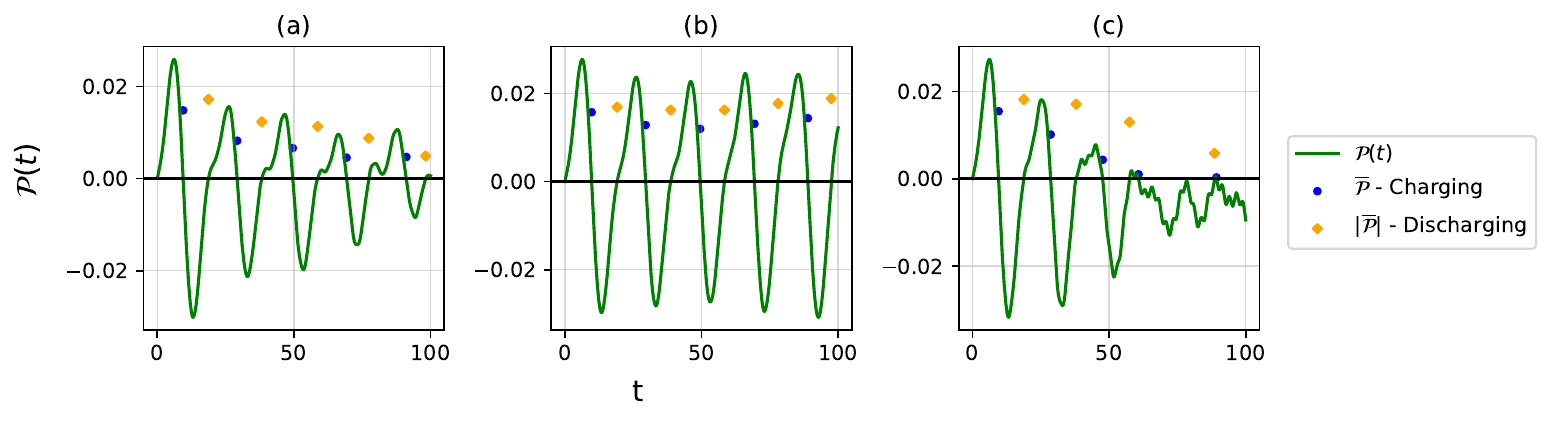}
    \caption{Variation of instantaneous charging power, average (dis-)charging power of battery spin with time for the two central spin charger-battery model for (a) $\lambda = 0.25$, (b) $\lambda = 0.5$, and (c) $\lambda = 1.0$. The parameters are: $\omega_C = 1.5$, $\omega_B = 1.25$, $\omega_{E_{C}}=0.7$, $\omega_{E_{B}}= 0.6$, $g_{CB}=0.05$, $g_{CE_{C}}=0.04$, $g_{BE_{B}}=0.02$, $T_{C}=0.5$, and $T_{B}=0.8$. Natural units are taken, where $\hbar = k_B = 1$.}
    \label{fig:Ins_AvgDisCharg_powerwith_time_2CentralSpin_model}
\end{figure*}
In Fig.~\ref{fig:Ergotropy_time_2CentralSpin_model_CB}, the time evolution of the ergotropy of the battery qubit for different values of magnetic field strengths $\lambda$ is depicted. The ergotropy exhibits an oscillatory behavior. Due to the interaction of the battery with the charger, the ergotropy initially increases, indicating charging of the battery. The ergotropy drops again due to its interaction with the environment, and this cycle continues. An interesting observation is made at the critical point $\lambda_c = 1.0$. At this point, the ergotropy decays rapidly at longer times, which indicates that the environment has a greater impact on its dissipation than its charging via the charger, and the battery almost completely loses the work that can be extracted from it, as ergotropy approaches zero in the long-time limit. The recharging of the battery here could be attributed to the non-Markovian system-bath interactions. At the critical point, this non-Markovian effect is highly suppressed, resulting in critical discharging. In Fig.~\ref{fig:Ergotropy_time_2CentralSpin_model_CB}(b) the ergotropy is plotted for a higher number of bath spins interacting with the charger and the battery at the critical coupling, where a consistent behavior of the ergotropy emerges, that is, the ergotropy decreases at the critical coupling and also with the increase in the bath spins.

Further, to investigate the behavior of the battery qubit near the critical point, we analyze the variation of ergotropy with the magnetic field strength, as shown in Fig.~\ref{fig:Ergotropy_lamda_2CentralSpin_model_CB}. At $t=20$, the ergotropy profile as a function of the magnetic-field strength is nearly stationary. For longer evolution times, for example, at $t = 40, 60$, and $80$,  local minima appear at several $\lambda$ values, while the ergotropy at large $\lambda$ values remains close to its earlier time values. However, at the critical point $\lambda_c = 1$, a sharp and pronounced minimum is observed, particularly at longer times, corresponding to the lowest ergotropy and a significantly larger drop than at other $\lambda$ values.

In Figs.~\ref{fig:Energyy_powerwith_lambda_2CentralSpin_model_CB} and \ref{fig:Energyy_powerwith_time_2CentralSpin_model_CB}, we discuss the energy and power of the battery with evolution time and magnetic field strength. The time evolution of the energy exhibits behavior similar to that of the ergotropy for different values of $\lambda$ shown in Fig.~\ref{fig:Ergotropy_time_2CentralSpin_model_CB}. The battery qubit loses energy through interactions with the bath and regains it back via the battery-charger coupling. As observed in Fig.~\ref{fig:Energyy_powerwith_time_2CentralSpin_model_CB}, the energy and power display qualitatively similar trends for $\lambda = 0.25, 0.5$, and $1.5$. When the system dissipates energy, power becomes negative, whereas energy absorption from the environment corresponds to positive power. Thus, energy and power remain consistent with one another, both showing oscillatory behavior. However, at the critical point, the amplitude of the energy oscillations decreases over longer evolution times, with the energy approaching its minimum attainable value for the battery qubit. Consequently, the power also exhibits damped oscillations. In particular, Fig.~\ref{fig:Energyy_powerwith_lambda_2CentralSpin_model_CB} illustrates that at $\lambda_c = 1$, during the evolution of energy and power, the energy attains its minimum value at a specific time, highlighting the distinct dynamical response near criticality.

Furthermore, the entanglement between the charger and the battery is studied to better understand the behavior of the battery at the critical point. To this end, we use concurrence $\mathcal{C}$ to explore the entanglement between the battery and charger qubit. This charger-battery setup forms a two-qubit bipartite system, in which case, the concurrence of the system is given by
\begin{equation}
	\mathcal{C} = \max\{0, \lambda_1 - \lambda_2 - \lambda_3 - \lambda_4\},
\end{equation}
where $\lambda_i's$ are the eigenvalue of the matrix $\sqrt{\sqrt{\rho_{CB}(t)}\tilde{\rho}_{CB}(t)\sqrt{\rho_{CB}(t)}}$ in decreasing order, and $\tilde{\rho}_{CB}(t) = (\sigma_y\otimes\sigma_y)\rho^*_{CB}(t)(\sigma_y\otimes\sigma_y)$. $\rho_{CB}(t)$ is the two-qubit system's density matrix given by Eq.~\eqref{rho_cbt}. The concurrence is calculated using the above formula, and is plotted in Fig.~\ref{fig:Concurrencewith_time_2CentralSpin_model_CB1}. The figure shows that at critical coupling, entanglement is higher at longer durations. This points out that a reason for less extractable work at the critical coupling could be attributed to a larger entanglement between the charger and the battery.

Up to this point, we have discussed that the battery qubit loses ergotropy due to system-bath interactions, while the battery-charger coupling helps it regain ergotropy, energy, and power. To further understand the charging behavior of the battery qubit, here we analyze the instantaneous charging power and the average (dis-)charging power in Fig.~\ref{fig:Ins_AvgDisCharg_powerwith_time_2CentralSpin_model}. Positive (negative) power indicates charging (discharging), corresponding to an increase (decrease) in ergotropy. In Fig.~\ref{fig:Ins_AvgDisCharg_powerwith_time_2CentralSpin_model}(a) and (b), for $\lambda = 0.25$ and $\lambda = 0.5$, the positive peaks of power cycles are slightly lower than the negative ones, resulting in an average charging power that is somewhat lower than the average discharging.
However, at the critical point $\lambda = 1$, shown in Fig~\ref{fig:Ins_AvgDisCharg_powerwith_time_2CentralSpin_model}(c), the positive peaks decay rapidly, while the negative part of the power persists. This behavior is consistent with the earlier observation that ergotropy decays faster at the critical point for long-time evolution. Consequently, compared with other values of $\lambda$, the average charging power is significantly decreased, while the average discharging power remains higher, thereby increasing the difference between the two. 

So far, we have investigated three distinct models of quantum batteries within an open quantum system framework. In the first two models, a two-qubit battery is considered to study the effect of inter-qubit interaction on battery performance. In the first model, we examined the role of inter-qubit interactions by comparing DM and Heisenberg XXX couplings in a two-qubit battery. In the second model, we investigated the effect of collective decoherence and temperature on extractable work. In both cases, the contribution of coherence to the battery performance was analyzed. Finally, to explore the role of entanglement and quantum criticality, we considered a charger–battery setup in which criticality is introduced through a spin-chain bath coupled to the charger. Together, these models provide a unified perspective on how interactions, environmental effects, coherence, entanglement, and critical behavior govern quantum battery dynamics.

\section{Conclusions}\label{conclusion}
In this work, three two-qubit models have been analyzed for quantum battery applications with distinct features. In the first model, the impact of the inter-qubit interactions, particularly the DM and Heisenberg XXX interaction, on the performance of the quantum battery was investigated.  The division of ergotropy into its (in)coherent parts for this model revealed that the incoherent ergotropy was responsible for the dominance of ergotropy in the XXX inter-qubit interaction case. This was further benchmarked by the charging power and average (dis-)charging power, highlighting that XXX inter-qubit interaction fares better at longer duration for work extraction applications. 

In the two-qubit collective decoherence model with a squeezed thermal bath, ergotropy dynamics were governed by interatomic distance and bath temperature. Collective decoherence led to slower decay of the ergotropy, with the coherent component dominating at long times, while independent decoherence resulted in a faster, monotonic loss of extractable work. Higher bath temperatures, in a squeezed thermal bath, accelerated ergotropy dissipation, whereas low-temperature environments helped in preserving it. Overall, collective effects and low temperatures were seen to enhance the longevity of ergotropy, highlighting their crucial role in maintaining the battery’s work-extraction capabilities.

To understand the effect of quantum criticality on battery performance, a two-qubit model composed of a charger and a battery qubit was considered. At the critical point $\lambda_c = 1$, ergotropy decayed rapidly, energy oscillations were damped, and power was predominantly negative, indicating that the environmental effects dominated over charging. The entanglement 
between the charger and the battery qubits was higher at the critical coupling and at longer durations, suggesting that higher entanglement between the charger and the battery may lead to lower work extraction from the battery.
Furthermore, at critical coupling, the instantaneous and average charging powers were significantly reduced, while discharging persisted, resulting in minimal extractable work. These results suggest that criticality strongly suppresses the long-time performance of the central spin quantum battery.

\bibliographystyle{apsrev4-2}
\bibliography{reference}

@book{sbbook,
author = {Banerjee, Subhashish},
title = {Open Quantum Systems: Dynamics of Nonclassical Evolution},
year = {2018},
isbn = {9789811331817},
publisher = {Springer Publishing Company, Incorporated},
edition = {1st}
}

@book{breuer2002book,
  title={The Theory of Open Quantum Systems},
  author={Breuer, H.P. and Petruccione, F.},
  isbn={9780198520634},
  lccn={2002075713},
  url={https://books.google.co.in/books?id=0Yx5VzaMYm8C},
  year={2002},
  publisher={Oxford University Press}
}

@book{weiss,
author = {Weiss, Ulrich},
title = {Quantum Dissipative Systems},
publisher = {WORLD SCIENTIFIC},
year = {2012},
doi = {10.1142/8334},
address = {},
edition   = {4th},
URL = {https://www.worldscientific.com/doi/abs/10.1142/8334},}

@article{gksl_master1,
    author = {Gorini, Vittorio and Kossakowski, Andrzej and Sudarshan, E. C. G.},
    title = "{Completely positive dynamical semigroups of N‐level systems}",
    journal = {Journal of Mathematical Physics},
    volume = {17},
    number = {5},
    pages = {821-825},
    year = {2008},
    month = {08},
    issn = {0022-2488},
    doi = {10.1063/1.522979},
    url = {https://doi.org/10.1063/1.522979}
}

@ARTICLE{Lindblad1976,
author={Lindblad, G.},
title={On the generators of quantum dynamical semigroups},
journal={Communications in Mathematical Physics},
year={1976},
month={Jun},
day={01},
volume={48},
number={2},
pages={119-130},
issn={1432-0916},
doi={10.1007/BF01608499},
url={https://doi.org/10.1007/BF01608499}
}

@article{Rivas_2014,
doi = {10.1088/0034-4885/77/9/094001},
url = {https://dx.doi.org/10.1088/0034-4885/77/9/094001},
year = {2014},
month = {aug},
publisher = {IOP Publishing},
volume = {77},
number = {9},
pages = {094001},
author = {Ángel Rivas and Susana F Huelga and Martin B Plenio},
title = {Quantum non-Markovianity: characterization, quantification and detection},
journal = {Reports on Progress in Physics}
}

@article{BLP_measure,
  title = {Measure for the Degree of Non-Markovian Behavior of Quantum Processes in Open Systems},
  author = {Breuer, Heinz-Peter and Laine, Elsi-Mari and Piilo, Jyrki},
  journal = {Phys. Rev. Lett.},
  volume = {103},
  issue = {21},
  pages = {210401},
  numpages = {4},
  year = {2009},
  month = {Nov},
  publisher = {American Physical Society},
  doi = {10.1103/PhysRevLett.103.210401},
  url = {https://link.aps.org/doi/10.1103/PhysRevLett.103.210401}
}

@article{de_vega_alonso,
  title = {Dynamics of non-Markovian open quantum systems},
  author = {de Vega, In\'es and Alonso, Daniel},
  journal = {Rev. Mod. Phys.},
  volume = {89},
  issue = {1},
  pages = {015001},
  numpages = {58},
  year = {2017},
  month = {Jan},
  publisher = {American Physical Society},
  doi = {10.1103/RevModPhys.89.015001},
 url = {https://link.aps.org/doi/10.1103/RevModPhys.89.015001}
}

@Article{Utagi2020,
author={Utagi, Shrikant
and Srikanth, R.
and Banerjee, Subhashish},
title={Temporal self-similarity of quantum dynamical maps as a concept of memorylessness},
journal={Scientific Reports},
year={2020},
month={Sep},
day={14},
volume={10},
number={1},
pages={15049},
issn={2045-2322},
doi={10.1038/s41598-020-72211-3},
url={https://doi.org/10.1038/s41598-020-72211-3}
}

@article{LI20181,
title = {Concepts of quantum non-Markovianity: A hierarchy},
journal = {Physics Reports},
volume = {759},
pages = {1-51},
year = {2018},
note = {Concepts of quantum non-Markovianity: A hierarchy},
issn = {0370-1573},
doi = {https://doi.org/10.1016/j.physrep.2018.07.001},
url = {https://www.sciencedirect.com/science/article/pii/S0370157318301601},
author = {Li Li and Michael J.W. Hall and Howard M. Wiseman}
}

@article{Chiranjib_2017,
  title = {Dynamics and thermodynamics of a central spin immersed in a spin bath},
  author = {Mukhopadhyay, Chiranjib and Bhattacharya, Samyadeb and Misra, Avijit and Pati, Arun Kumar},
  journal = {Phys. Rev. A},
  volume = {96},
  issue = {5},
  pages = {052125},
  numpages = {13},
  year = {2017},
  month = {Nov},
  publisher = {American Physical Society},
  doi = {10.1103/PhysRevA.96.052125},
  url = {https://link.aps.org/doi/10.1103/PhysRevA.96.052125}
}

@article{CHRUSCINSKI20221,
title = {Dynamical maps beyond Markovian regime},
journal = {Physics Reports},
volume = {992},
pages = {1-85},
year = {2022},
note = {Dynamical maps beyond Markovian regime},
issn = {0370-1573},
doi = {https://doi.org/10.1016/j.physrep.2022.09.003},
url = {https://www.sciencedirect.com/science/article/pii/S0370157322003428},
author = {Dariusz Chruściński}
}

@article{CALDEIRA1983_tunneling,
title = {Quantum tunnelling in a dissipative system},
journal = {Annals of Physics},
volume = {149},
number = {2},
pages = {374-456},
year = {1983},
issn = {0003-4916},
doi = {https://doi.org/10.1016/0003-4916(83)90202-6},
url = {https://www.sciencedirect.com/science/article/pii/0003491683902026},
author = {A.O Caldeira and A.J Leggett}
}

@article{Legget_dissipative,
  title = {Dynamics of the dissipative two-state system},
  author = {Leggett, A. J. and Chakravarty, S. and Dorsey, A. T. and Fisher, Matthew P. A. and Garg, Anupam and Zwerger, W.},
  journal = {Rev. Mod. Phys.},
  volume = {59},
  issue = {1},
  pages = {1--85},
  numpages = {0},
  year = {1987},
  month = {Jan},
  publisher = {American Physical Society},
  doi = {10.1103/RevModPhys.59.1},
  url = {https://link.aps.org/doi/10.1103/RevModPhys.59.1}
}

@article{Burgarth2004,
  title = {Non-Markovian dynamics in a spin star system: Exact solution and approximation techniques},
  author = {Breuer, Heinz-Peter and Burgarth, Daniel and Petruccione, Francesco},
  journal = {Phys. Rev. B},
  volume = {70},
  issue = {4},
  pages = {045323},
  numpages = {10},
  year = {2004},
  month = {Jul},
  publisher = {American Physical Society},
  doi = {10.1103/PhysRevB.70.045323},
  url = {https://link.aps.org/doi/10.1103/PhysRevB.70.045323}
}

@article{BANERJEE2010,
title = {Dynamics of entanglement in two-qubit open system interacting with a squeezed thermal bath via dissipative interaction},
journal = {Annals of Physics},
volume = {325},
number = {4},
pages = {816-834},
year = {2010},
issn = {0003-4916},
doi = {https://doi.org/10.1016/j.aop.2010.01.003},
url = {https://www.sciencedirect.com/science/article/pii/S0003491610000060},
author = {Subhashish Banerjee and V. Ravishankar and R. Srikanth}
}

@book{gemmer2004quantum,
  title={Quantum Thermodynamics: Emergence of Thermodynamic Behavior Within Composite Quantum Systems},
  author={Gemmer, J. and Michel, M. and Mahler, G.},
  isbn={9783540229117},
  lccn={2004110894},
  series={Lecture Notes in Physics},
  url={https://books.google.co.in/books?id=MqDUIvOCIgoC},
  year={2004},
  publisher={Springer Berlin Heidelberg}
}

@book{binder2019thermodynamics,
  title={Thermodynamics in the Quantum Regime: Fundamental Aspects and New Directions},
  author={Binder, F. and Correa, L.A. and Gogolin, C. and Anders, J. and Adesso, G.},
  isbn={9783319990453},
  series={Fundamental Theories of Physics},
  url={https://books.google.co.in/books?id=IQlouQEACAAJ},
  year={2019},
  publisher={Springer International Publishing}
}

@article{sai_anders_book,
author = {Sai Vinjanampathy and Janet Anders},
title = {Quantum thermodynamics},
journal = {Contemporary Physics},
volume = {57},
number = {4},
pages = {545-579},
year  = {2016},
publisher = {Taylor & Francis},
doi = {10.1080/00107514.2016.1201896},
URL = {https://doi.org/10.1080/00107514.2016.1201896},
eprint = {https://doi.org/10.1080/00107514.2016.1201896}
}

@book{deffner2019quantum,
  title={Quantum Thermodynamics: An Introduction to the Thermodynamics of Quantum Information},
  author={Deffner, S. and Campbell, S.E. and Institute of Physics (Gran Bretanya) and Morgan \& Claypool Publishers},
  isbn={9781643276564},
  series={IOP (Series).: Release 6},
  url={https://books.google.co.in/books?id=-IwbyAEACAAJ},
  year={2019},
  publisher={Morgan \& Claypool Publishers}
}

@article{Seifert_2012,
doi = {10.1088/0034-4885/75/12/126001},
url = {https://dx.doi.org/10.1088/0034-4885/75/12/126001},
year = {2012},
month = {nov},
publisher = {IOP Publishing},
volume = {75},
number = {12},
pages = {126001},
author = {Udo Seifert},
title = {Stochastic thermodynamics, fluctuation theorems and molecular machines},
journal = {Reports on Progress in Physics}
}

@article{Hanggi_talkner,
  title = {Colloquium: Statistical mechanics and thermodynamics at strong coupling: Quantum and classical},
  author = {Talkner, Peter and H\"anggi, Peter},
  journal = {Rev. Mod. Phys.},
  volume = {92},
  issue = {4},
  pages = {041002},
  numpages = {26},
  year = {2020},
  month = {Oct},
  publisher = {American Physical Society},
  doi = {10.1103/RevModPhys.92.041002},
  url = {https://link.aps.org/doi/10.1103/RevModPhys.92.041002}
}

@book{sekimoto2010stochastic,
  title={Stochastic Energetics},
  author={Sekimoto, K.},
  isbn={9783642054112},
  lccn={2009943129},
  series={Lecture Notes in Physics},
  url={https://books.google.co.in/books?id=8Fq7BQAAQBAJ},
  year={2010},
  publisher={Springer Berlin Heidelberg}
}

@Inbook{Alicki2018_Kosloff,
author="Alicki, Robert
and Kosloff, Ronnie",
editor="Binder, Felix
and Correa, Luis A.
and Gogolin, Christian
and Anders, Janet
and Adesso, Gerardo",
title="Introduction to Quantum Thermodynamics: History and Prospects",
bookTitle="Thermodynamics in the Quantum Regime: Fundamental Aspects and New Directions",
year="2018",
publisher="Springer International Publishing",
address="Cham",
pages="1--33",
doi="10.1007/978-3-319-99046-0_1",
url="https://doi.org/10.1007/978-3-319-99046-0_1"
}

@article{Allahverdyan_2004,
doi = {10.1209/epl/i2004-10101-2},
url = {https://dx.doi.org/10.1209/epl/i2004-10101-2},
year = {2004},
month = {aug},
publisher = {},
volume = {67},
number = {4},
pages = {565},
author = {A. E. Allahverdyan and  R. Balian and  Th. M. Nieuwenhuizen},
title = {Maximal work extraction from finite quantum systems},
journal = {Europhysics Letters}
}

@article{cakmak1,
  title = {Ergotropy from coherences in an open quantum system},
  author = {Bar{\i}{\c{s}} {\c{C}}akmak},
  journal = {Phys. Rev. E},
  volume = {102},
  issue = {4},
  pages = {042111},
  numpages = {11},
  year = {2020},
  month = {Oct},
  publisher = {American Physical Society},
  doi = {10.1103/PhysRevE.102.042111},
  url = {https://link.aps.org/doi/10.1103/PhysRevE.102.042111}
}

@article{coherent_ergo1,
  title = {Quantum Coherence and Ergotropy},
  author = {Francica, G. and Binder, F. C. and Guarnieri, G. and Mitchison, M. T. and Goold, J. and Plastina, F.},
  journal = {Phys. Rev. Lett.},
  volume = {125},
  issue = {18},
  pages = {180603},
  numpages = {8},
  year = {2020},
  month = {Oct},
  publisher = {American Physical Society},
  doi = {10.1103/PhysRevLett.125.180603},
  url = {https://link.aps.org/doi/10.1103/PhysRevLett.125.180603}
}

@article{thomas_heat_engine,
  title = {Thermodynamics of non-Markovian reservoirs and heat engines},
  author = {Thomas, George and Siddharth, Nana and Banerjee, Subhashish and Ghosh, Sibasish},
  journal = {Phys. Rev. E},
  volume = {97},
  issue = {6},
  pages = {062108},
  numpages = {8},
  year = {2018},
  month = {Jun},
  publisher = {American Physical Society},
  doi = {10.1103/PhysRevE.97.062108},
  url = {https://link.aps.org/doi/10.1103/PhysRevE.97.062108}
}

@article{KUMAR2023128832,
title = {Thermodynamics of one and two-qubit nonequilibrium heat engines running between squeezed thermal reservoirs},
journal = {Physica A: Statistical Mechanics and its Applications},
volume = {623},
pages = {128832},
year = {2023},
issn = {0378-4371},
doi = {https://doi.org/10.1016/j.physa.2023.128832},
url = {https://www.sciencedirect.com/science/article/pii/S0378437123003874},
author = {Ashutosh Kumar and Sourabh Lahiri and Trilochan Bagarti and Subhashish Banerjee}
}

@article{Cresser_2014,
  title = {Canonical form of master equations and characterization of non-Markovianity},
  author = {Hall, Michael J. W. and Cresser, James D. and Li, Li and Andersson, Erika},
  journal = {Phys. Rev. A},
  volume = {89},
  issue = {4},
  pages = {042120},
  numpages = {11},
  year = {2014},
  month = {Apr},
  publisher = {American Physical Society},
  doi = {10.1103/PhysRevA.89.042120},
  url = {https://link.aps.org/doi/10.1103/PhysRevA.89.042120}
}

@article{Alicki_Fannes_QB,
  title = {Entanglement boost for extractable work from ensembles of quantum batteries},
  author = {Alicki, Robert and Fannes, Mark},
  journal = {Phys. Rev. E},
  volume = {87},
  issue = {4},
  pages = {042123},
  numpages = {4},
  year = {2013},
  month = {Apr},
  publisher = {American Physical Society},
  doi = {10.1103/PhysRevE.87.042123},
  url = {https://link.aps.org/doi/10.1103/PhysRevE.87.042123}
}

@article{Binder_2015,
doi = {10.1088/1367-2630/17/7/075015},
url = {https://dx.doi.org/10.1088/1367-2630/17/7/075015},
year = {2015},
month = {jul},
publisher = {IOP Publishing},
volume = {17},
number = {7},
pages = {075015},
author = {Felix C Binder and Sai Vinjanampathy and Kavan Modi and John Goold},
title = {Quantacell: powerful charging of quantum batteries},
journal = {New Journal of Physics},
}

@misc{campaioli2018quantum,
      title={Quantum Batteries - Review Chapter}, 
      author={Francesco Campaioli and Felix A. Pollock and Sai Vinjanampathy},
      year={2018},
      eprint={1805.05507},
      archivePrefix={arXiv},
      primaryClass={quant-ph}
}

@article{Campaioli_2017,
  title = {Enhancing the Charging Power of Quantum Batteries},
  author = {Campaioli, Francesco and Pollock, Felix A. and Binder, Felix C. and C\'eleri, Lucas and Goold, John and Vinjanampathy, Sai and Modi, Kavan},
  journal = {Phys. Rev. Lett.},
  volume = {118},
  issue = {15},
  pages = {150601},
  numpages = {6},
  year = {2017},
  month = {Apr},
  publisher = {American Physical Society},
  doi = {10.1103/PhysRevLett.118.150601},
  url = {https://link.aps.org/doi/10.1103/PhysRevLett.118.150601}
}

@article{Bhattacharya2021,
  title = {Revisiting the Quantum Open System Dynamics of Central Spin Model},
  volume = {10},
  ISSN = {1314-7374},
  url = {http://dx.doi.org/10.12743/quanta.v10i1.162},
  DOI = {10.12743/quanta.v10i1.162},
  number = {1},
  journal = {Quanta},
  publisher = {Quanta},
  author = {Bhattacharya,  Samyadeb and Banerjee,  Subhashish},
  year = {2021},
  month = dec,
  pages = {55–64}
}

@article{bhanja2023,
  title = {Impact of non-Markovian quantum Brownian motion on quantum batteries},
  author = {Bhanja, Gourab and Tiwari, Devvrat and Banerjee, Subhashish},
  journal = {Phys. Rev. A},
  volume = {109},
  issue = {1},
  pages = {012224},
  numpages = {9},
  year = {2024},
  month = {Jan},
  publisher = {American Physical Society},
  doi = {10.1103/PhysRevA.109.012224},
  url = {https://link.aps.org/doi/10.1103/PhysRevA.109.012224}
}

@article{Ferraro_2018,
  title = {High-Power Collective Charging of a Solid-State Quantum Battery},
  author = {Ferraro, Dario and Campisi, Michele and Andolina, Gian Marcello and Pellegrini, Vittorio and Polini, Marco},
  journal = {Phys. Rev. Lett.},
  volume = {120},
  issue = {11},
  pages = {117702},
  numpages = {6},
  year = {2018},
  month = {Mar},
  publisher = {American Physical Society},
  doi = {10.1103/PhysRevLett.120.117702},
  url = {https://link.aps.org/doi/10.1103/PhysRevLett.120.117702}
}

@article{QB_open_system,
  title = {Charger-mediated energy transfer for quantum batteries: An open-system approach},
  author = {Farina, Donato and Andolina, Gian Marcello and Mari, Andrea and Polini, Marco and Giovannetti, Vittorio},
  journal = {Phys. Rev. B},
  volume = {99},
  issue = {3},
  pages = {035421},
  numpages = {15},
  year = {2019},
  month = {Jan},
  publisher = {American Physical Society},
  doi = {10.1103/PhysRevB.99.035421},
  url = {https://link.aps.org/doi/10.1103/PhysRevB.99.035421}
}

@article{Kamin_2020,
doi = {10.1088/1367-2630/ab9ee2},
url = {https://dx.doi.org/10.1088/1367-2630/ab9ee2},
year = {2020},
month = {aug},
publisher = {IOP Publishing},
volume = {22},
number = {8},
pages = {083007},
author = {F H Kamin and F T Tabesh and S Salimi and F Kheirandish and Alan C Santos},
title = {Non-Markovian effects on charging and self-discharging process of quantum batteries},
journal = {New Journal of Physics}
}

@article{PhysRevA.106.032435,
  title = {Dynamics of two central spins immersed in spin baths},
  author = {Tiwari, Devvrat and Datta, Shounak and Bhattacharya, Samyadeb and Banerjee, Subhashish},
  journal = {Phys. Rev. A},
  volume = {106},
  issue = {3},
  pages = {032435},
  numpages = {15},
  year = {2022},
  month = {Sep},
  publisher = {American Physical Society},
  doi = {10.1103/PhysRevA.106.032435},
  url = {https://link.aps.org/doi/10.1103/PhysRevA.106.032435}
}

@ARTICLE{Devvrat_impact,
AUTHOR={Tiwari, Devvrat and Banerjee, Subhashish},   	 
TITLE={Impact of non-Markovian evolution on characterizations of quantum thermodynamics},      	
JOURNAL={Frontiers in Quantum Science and Technology},      	
VOLUME={2},           	
YEAR={2023},
pages = {1207552},
URL={https://www.frontiersin.org/articles/10.3389/frqst.2023.1207552},       	
DOI={10.3389/frqst.2023.1207552},      	
ISSN={2813-2181}
}

@article{devvrat_central_spin_2,
author = {Tiwari, Devvrat and Paulson, Kavalambramalil G. and Banerjee, Subhashish},
title = {Quantum Correlations and Speed Limit of Central Spin Systems},
journal = {Annalen der Physik},
volume = {535},
number = {2},
pages = {2200452},
doi = {https://doi.org/10.1002/andp.202200452},
url = {https://onlinelibrary.wiley.com/doi/abs/10.1002/andp.202200452},
year = {2023}
}

@article{Wei2016,
author={Wei, Yong-Bonand Zou, Jian and Wang, Zhao-Ming and Shao, Bin},
title={Quantum speed limit and a signal of quantum criticality},
journal={Scientific Reports},
year={2016},
month={Jan},
day={19},
volume={6},
number={1},
pages={19308},
issn={2045-2322},
doi={10.1038/srep19308},
url={https://doi.org/10.1038/srep19308}
}

@book{sachdev2011quantum,
  title={Quantum Phase Transitions},
  author={Sachdev, S.},
  isbn={9781139500210},
  url={https://books.google.co.in/books?id=F3IkpxwpqSgC},
  year={2011},
  publisher={Cambridge University Press, Cambridge UK}
}

@article{yadav2025thermo,
author = {Yadav, Mahima and Tiwari, Devvrat and Banerjee, Subhashish},
title = {(Thermo-)Dynamics of the Spin-Boson Model in the Weak Coupling Regime: Application as a Quantum Battery},
journal = {Advanced Quantum Technologies},
volume = {n/a},
number = {n/a},
pages = {e00333},
doi = {https://doi.org/10.1002/qute.202500333},
url = {https://advanced.onlinelibrary.wiley.com/doi/abs/10.1002/qute.202500333}
}

@article{landi_review_entropy_production,
  title = {Irreversible entropy production: From classical to quantum},
  author = {Landi, Gabriel T. and Paternostro, Mauro},
  journal = {Rev. Mod. Phys.},
  volume = {93},
  issue = {3},
  pages = {035008},
  numpages = {58},
  year = {2021},
  month = {Sep},
  publisher = {American Physical Society},
  doi = {10.1103/RevModPhys.93.035008},
  url = {https://link.aps.org/doi/10.1103/RevModPhys.93.035008}
}

@article{SB2023_thermalization,
title = {Thermalization in quenched open quantum cosmology},
journal = {Nuclear Physics B},
volume = {996},
pages = {116368},
year = {2023},
issn = {0550-3213},
doi = {https://doi.org/10.1016/j.nuclphysb.2023.116368},
url = {https://www.sciencedirect.com/science/article/pii/S0550321323002973},
author = {Subhashish Banerjee and Sayantan Choudhury and Satyaki Chowdhury and Johannes Knaute and Sudhakar Panda and K. Shirish}
}

@article{devvrat_strong,
    author = {Tiwari, Devvrat and Bose, Baibhab and Banerjee, Subhashish},
    title = {Strong coupling non-Markovian quantum thermodynamics of a finite-bath system},
    journal = {The Journal of Chemical Physics},
    volume = {162},
    number = {11},
    pages = {114104},
    year = {2025},
    month = {03},
    issn = {0021-9606},
    doi = {10.1063/5.0254029},
    url = {https://doi.org/10.1063/5.0254029}
}

@Article{devvrat-neha_hmf,
author={Pathania, Neha
and Tiwari, Devvrat
and Banerjee, Subhashish},
title={Quantum thermodynamics of open quantum systems: nature of thermal fluctuations},
journal={Quantum Information Processing},
year={2025},
month={Aug},
day={28},
volume={24},
number={9},
pages={290},
issn={1573-1332},
doi={10.1007/s11128-025-04903-6},
url={https://doi.org/10.1007/s11128-025-04903-6}
}

@article{Moriya,
  title = {Anisotropic Superexchange Interaction and Weak Ferromagnetism},
  author = {Moriya, T\^oru},
  journal = {Phys. Rev.},
  volume = {120},
  issue = {1},
  pages = {91--98},
  numpages = {0},
  year = {1960},
  month = {Oct},
  publisher = {American Physical Society},
  doi = {10.1103/PhysRev.120.91},
  url = {https://link.aps.org/doi/10.1103/PhysRev.120.91}
}

@article{DZYALOSHINSKY1958,
title = {A thermodynamic theory of “weak” ferromagnetism of antiferromagnetics},
journal = {Journal of Physics and Chemistry of Solids},
volume = {4},
number = {4},
pages = {241-255},
year = {1958},
issn = {0022-3697},
doi = {https://doi.org/10.1016/0022-3697(58)90076-3},
url = {https://www.sciencedirect.com/science/article/pii/0022369758900763},
author = {I. Dzyaloshinsky}
}

@article{spin_chain_primer_nepomechie,
author = {NEPOMECHIE, RAFAEL I.},
title = {A SPIN CHAIN PRIMER},
journal = {International Journal of Modern Physics B},
volume = {13},
number = {24n25},
pages = {2973-2985},
year = {1999},
doi = {10.1142/S0217979299002800},
URL = {https://doi.org/10.1142/S0217979299002800}
}

@book{Mattis1981,
  title     = "The theory of magnetism I",
  author    = "Mattis, Daniel C",
  publisher = "Springer",
  series    = "Springer Series in Solid-State Sciences",
  edition   =  1981,
  month     =  aug,
  year      =  1981,
  address   = "Berlin, Germany"
}

@Article{Heisenberg1928,
author={Heisenberg, W.},
title={Zur Theorie des Ferromagnetismus},
journal={Zeitschrift f{\"u}r Physik},
year={1928},
month={Sep},
day={01},
volume={49},
number={9},
pages={619-636},
issn={0044-3328},
doi={10.1007/BF01328601},
url={https://doi.org/10.1007/BF01328601}
}

@Article{Bethe1931,
author={Bethe, H.},
title={Zur Theorie der Metalle},
journal={Zeitschrift f{\"u}r Physik},
year={1931},
month={Mar},
day={01},
volume={71},
number={3},
pages={205-226},
issn={0044-3328},
doi={10.1007/BF01341708},
url={https://doi.org/10.1007/BF01341708}
}

@article{Gour_2022,
  title = {Role of Quantum Coherence in Thermodynamics},
  author = {Gour, Gilad},
  journal = {PRX Quantum},
  volume = {3},
  issue = {4},
  pages = {040323},
  numpages = {23},
  year = {2022},
  month = {Nov},
  publisher = {American Physical Society},
  doi = {10.1103/PRXQuantum.3.040323},
  url = {https://link.aps.org/doi/10.1103/PRXQuantum.3.040323}
}

@article{Alicki_1979,
doi = {10.1088/0305-4470/12/5/007},
url = {https://doi.org/10.1088/0305-4470/12/5/007},
year = {1979},
month = {may},
publisher = {},
volume = {12},
number = {5},
pages = {L103},
author = {R Alicki},
title = {The quantum open system as a model of the heat engine},
journal = {Journal of Physics A: Mathematical and General}
}

@article{prasanna_heat_engine,
  title = {Quantum Statistical Enhancement of the Collective Performance of Multiple Bosonic Engines},
  author = {Watanabe, Gentaro and Venkatesh, B. Prasanna and Talkner, Peter and Hwang, Myung-Joong and del Campo, Adolfo},
  journal = {Phys. Rev. Lett.},
  volume = {124},
  issue = {21},
  pages = {210603},
  numpages = {6},
  year = {2020},
  month = {May},
  publisher = {American Physical Society},
  doi = {10.1103/PhysRevLett.124.210603},
  url = {https://link.aps.org/doi/10.1103/PhysRevLett.124.210603}
}

@article{devvrat_circuit,
  title = {Quantum Thermal Analogs of Electric Circuits: A Universal Approach},
  author = {Tiwari, Devvrat and Bhattacharya, Samyadeb and Banerjee, Subhashish},
  journal = {Phys. Rev. Lett.},
  volume = {135},
  issue = {2},
  pages = {020404},
  numpages = {6},
  year = {2025},
  month = {Jul},
  publisher = {American Physical Society},
  doi = {10.1103/5x8m-bhgd},
  url = {https://link.aps.org/doi/10.1103/5x8m-bhgd}
}

@article{diode_2017,
  title = {Quantum thermal diode based on two interacting spinlike systems under different excitations},
  author = {Ordonez-Miranda, Jose and Ezzahri, Youn\`es and Joulain, Karl},
  journal = {Phys. Rev. E},
  volume = {95},
  issue = {2},
  pages = {022128},
  numpages = {7},
  year = {2017},
  month = {Feb},
  publisher = {American Physical Society},
  doi = {10.1103/PhysRevE.95.022128},
  url = {https://link.aps.org/doi/10.1103/PhysRevE.95.022128}
}

@article{transistor_2016,
  title = {Quantum Thermal Transistor},
  author = {Joulain, Karl and Drevillon, J\'er\'emie and Ezzahri, Youn\`es and Ordonez-Miranda, Jose},
  journal = {Phys. Rev. Lett.},
  volume = {116},
  issue = {20},
  pages = {200601},
  numpages = {5},
  year = {2016},
  month = {May},
  publisher = {American Physical Society},
  doi = {10.1103/PhysRevLett.116.200601},
  url = {https://link.aps.org/doi/10.1103/PhysRevLett.116.200601}
}

@article{floquet_transistor_2022,
  title = {Floquet quantum thermal transistor},
  author = {Gupt, Nikhil and Bhattacharyya, Srijan and Das, Bikash and Datta, Subhadeep and Mukherjee, Victor and Ghosh, Arnab},
  journal = {Phys. Rev. E},
  volume = {106},
  issue = {2},
  pages = {024110},
  numpages = {14},
  year = {2022},
  month = {Aug},
  publisher = {American Physical Society},
  doi = {10.1103/PhysRevE.106.024110},
  url = {https://link.aps.org/doi/10.1103/PhysRevE.106.024110}
}

@article{wheatstone_bridge_2022,
  title = {Quantum Wheatstone Bridge},
  author = {Poulsen, Kasper and Santos, Alan C. and Zinner, Nikolaj T.},
  journal = {Phys. Rev. Lett.},
  volume = {128},
  issue = {24},
  pages = {240401},
  numpages = {6},
  year = {2022},
  month = {Jun},
  publisher = {American Physical Society},
  doi = {10.1103/PhysRevLett.128.240401},
  url = {https://link.aps.org/doi/10.1103/PhysRevLett.128.240401}
}

@article{QB_colloquium,
  title = {Colloquium: Quantum batteries},
  author = {Campaioli, Francesco and Gherardini, Stefano and Quach, James Q. and Polini, Marco and Andolina, Gian Marcello},
  journal = {Rev. Mod. Phys.},
  volume = {96},
  issue = {3},
  pages = {031001},
  numpages = {30},
  year = {2024},
  month = {Jul},
  publisher = {American Physical Society},
  doi = {10.1103/RevModPhys.96.031001},
  url = {https://link.aps.org/doi/10.1103/RevModPhys.96.031001}
}

@article{QB_andolina_2019,
  title = {Extractable Work, the Role of Correlations, and Asymptotic Freedom in Quantum Batteries},
  author = {Andolina, Gian Marcello and Keck, Maximilian and Mari, Andrea and Campisi, Michele and Giovannetti, Vittorio and Polini, Marco},
  journal = {Phys. Rev. Lett.},
  volume = {122},
  issue = {4},
  pages = {047702},
  numpages = {5},
  year = {2019},
  month = {Feb},
  publisher = {American Physical Society},
  doi = {10.1103/PhysRevLett.122.047702},
  url = {https://link.aps.org/doi/10.1103/PhysRevLett.122.047702}
}

@article{QB_advantage_3,
  title = {Reliable quantum advantage in quantum battery charging},
  author = {Rinaldi, Davide and Filip, Radim and Gerace, Dario and Guarnieri, Giacomo},
  journal = {Phys. Rev. A},
  volume = {112},
  issue = {1},
  pages = {012205},
  numpages = {11},
  year = {2025},
  month = {Jul},
  publisher = {American Physical Society},
  doi = {10.1103/6kwv-z6fx},
  url = {https://link.aps.org/doi/10.1103/6kwv-z6fx}
}

@article{QB_advantage_1,
  title = {Genuine Quantum Advantage in Anharmonic Bosonic Quantum Batteries},
  author = {Andolina, Gian Marcello and Stanzione, Vittoria and Giovannetti, Vittorio and Polini, Marco},
  journal = {Phys. Rev. Lett.},
  volume = {134},
  issue = {24},
  pages = {240403},
  numpages = {8},
  year = {2025},
  month = {Jun},
  publisher = {American Physical Society},
  doi = {10.1103/kzvn-dj7v},
  url = {https://link.aps.org/doi/10.1103/kzvn-dj7v}
}

@Article{Shastri2025,
author={Shastri, Rahul
and Jiang, Chao
and Xu, Guo-Hua
and Prasanna Venkatesh, B.
and Watanabe, Gentaro},
title={Dephasing enabled fast charging of quantum batteries},
journal={npj Quantum Information},
year={2025},
month={Jan},
day={19},
volume={11},
number={1},
pages={9},
issn={2056-6387},
doi={10.1038/s41534-025-00959-5},
url={https://doi.org/10.1038/s41534-025-00959-5}
}

@article{Extended_Dicke_battery,
  title = {Extended Dicke quantum battery with interatomic interactions and driving field},
  author = {Dou, Fu-Quan and Lu, You-Qi and Wang, Yuan-Jin and Sun, Jian-An},
  journal = {Phys. Rev. B},
  volume = {105},
  issue = {11},
  pages = {115405},
  numpages = {13},
  year = {2022},
  month = {Mar},
  publisher = {American Physical Society},
  doi = {10.1103/PhysRevB.105.115405},
  url = {https://link.aps.org/doi/10.1103/PhysRevB.105.115405}
}

@article{spin_chain_quantum_battery,
  title = {Spin-chain model of a many-body quantum battery},
  author = {Le, Thao P. and Levinsen, Jesper and Modi, Kavan and Parish, Meera M. and Pollock, Felix A.},
  journal = {Phys. Rev. A},
  volume = {97},
  issue = {2},
  pages = {022106},
  numpages = {9},
  year = {2018},
  month = {Feb},
  publisher = {American Physical Society},
  doi = {10.1103/PhysRevA.97.022106},
  url = {https://link.aps.org/doi/10.1103/PhysRevA.97.022106}
}

@article{central_spin_quantum_battery,
  title = {Entanglement and work extraction in the central-spin quantum battery},
  author = {Liu, Jia-Xuan and Shi, Hai-Long and Shi, Yun-Hao and Wang, Xiao-Hui and Yang, Wen-Li},
  journal = {Phys. Rev. B},
  volume = {104},
  issue = {24},
  pages = {245418},
  numpages = {8},
  year = {2021},
  month = {Dec},
  publisher = {American Physical Society},
  doi = {10.1103/PhysRevB.104.245418},
  url = {https://link.aps.org/doi/10.1103/PhysRevB.104.245418}
}

@article{SYK_battery,
  title = {Quantum Advantage in the Charging Process of Sachdev-Ye-Kitaev Batteries},
  author = {Rossini, Davide and Andolina, Gian Marcello and Rosa, Dario and Carrega, Matteo and Polini, Marco},
  journal = {Phys. Rev. Lett.},
  volume = {125},
  issue = {23},
  pages = {236402},
  numpages = {6},
  year = {2020},
  month = {Dec},
  publisher = {American Physical Society},
  doi = {10.1103/PhysRevLett.125.236402},
  url = {https://link.aps.org/doi/10.1103/PhysRevLett.125.236402}
}

@article{self_discharge_mitigated_QB,
  title = {Self-Discharging Mitigated Quantum Battery},
  author = {Song, Wan-Lu and Wang, Ji-Ling and Zhou, Bin and Yang, Wan-Li and An, Jun-Hong},
  journal = {Phys. Rev. Lett.},
  volume = {135},
  issue = {2},
  pages = {020405},
  numpages = {8},
  year = {2025},
  month = {Jul},
  publisher = {American Physical Society},
  doi = {10.1103/d9k1-75d4},
  url = {https://link.aps.org/doi/10.1103/d9k1-75d4}
}

@article{resonator-qutrit_QB,
  title = {Resonator-qutrit quantum battery},
  author = {Yang, Fang-Mei and Dou, Fu-Quan},
  journal = {Phys. Rev. A},
  volume = {109},
  issue = {6},
  pages = {062432},
  numpages = {9},
  year = {2024},
  month = {Jun},
  publisher = {American Physical Society},
  doi = {10.1103/PhysRevA.109.062432},
  url = {https://link.aps.org/doi/10.1103/PhysRevA.109.062432}
}

@article{Rosen-Zener_QB,
  title = {Analytically solvable many-body Rosen-Zener quantum battery},
  author = {Guo, Wei-Xi and Yang, Fang-Mei and Dou, Fu-Quan},
  journal = {Phys. Rev. A},
  volume = {109},
  issue = {3},
  pages = {032201},
  numpages = {9},
  year = {2024},
  month = {Mar},
  publisher = {American Physical Society},
  doi = {10.1103/PhysRevA.109.032201},
  url = {https://link.aps.org/doi/10.1103/PhysRevA.109.032201}
}

@Article{Pusz1978,
author={Pusz, W.
and Woronowicz, S. L.},
title={Passive states and KMS states for general quantum systems},
journal={Communications in Mathematical Physics},
year={1978},
month={Oct},
day={01},
volume={58},
number={3},
pages={273-290},
issn={1432-0916},
doi={10.1007/BF01614224},
url={https://doi.org/10.1007/BF01614224}
}

@article{Morrone_2023,
doi = {10.1088/2058-9565/accca4},
url = {https://doi.org/10.1088/2058-9565/accca4},
year = {2023},
month = {may},
publisher = {IOP Publishing},
volume = {8},
number = {3},
pages = {035007},
author = {Morrone, Daniele and Rossi, Matteo A C and Smirne, Andrea and Genoni, Marco G},
title = {Charging a quantum battery in a non-Markovian environment: a collisional model approach},
journal = {Quantum Science and Technology}
}

@article{prokofev_stamp_2000,
doi = {10.1088/0034-4885/63/4/204},
url = {https://dx.doi.org/10.1088/0034-4885/63/4/204},
year = {2000},
month = {apr},
publisher = {},
volume = {63},
number = {4},
pages = {669},
author = {N V Prokof'ev and  P C E Stamp},
title = {Theory 
of the spin bath},
journal = {Reports on Progress in Physics}
}

@Article{Prokofaev1996,
author={Prokof'ev, N. V.
and Stamp, P. C. E.},
title={Quantum relaxation of magnetisation in magnetic particles},
journal={Journal of Low Temperature Physics},
year={1996},
month={Aug},
day={01},
volume={104},
number={3},
pages={143-209},
issn={1573-7357},
doi={10.1007/BF00754094},
url={https://doi.org/10.1007/BF00754094}
}

@article{Haase_2018,
doi = {10.1088/1367-2630/aab67f},
url = {https://dx.doi.org/10.1088/1367-2630/aab67f},
year = {2018},
month = {may},
publisher = {IOP Publishing},
volume = {20},
number = {5},
pages = {053009},
author = {Haase, J F and Smirne, A and Kołodyński, J and Demkowicz-Dobrzański, R and Huelga, S F},
title = {Fundamental limits to frequency estimation: a comprehensive microscopic perspective},
journal = {New Journal of Physics}
}

@article{Hanson_2006,
  title = {Polarization and Readout of Coupled Single Spins in Diamond},
  author = {Hanson, R. and Mendoza, F. M. and Epstein, R. J. and Awschalom, D. D.},
  journal = {Phys. Rev. Lett.},
  volume = {97},
  issue = {8},
  pages = {087601},
  numpages = {4},
  year = {2006},
  month = {Aug},
  publisher = {American Physical Society},
  doi = {10.1103/PhysRevLett.97.087601},
  url = {https://link.aps.org/doi/10.1103/PhysRevLett.97.087601}
}

@article{trapped_ions,
  title = {Quantum Computations with Cold Trapped Ions},
  author = {Cirac, J. I. and Zoller, P.},
  journal = {Phys. Rev. Lett.},
  volume = {74},
  issue = {20},
  pages = {4091--4094},
  numpages = {0},
  year = {1995},
  month = {May},
  publisher = {American Physical Society},
  doi = {10.1103/PhysRevLett.74.4091},
  url = {https://link.aps.org/doi/10.1103/PhysRevLett.74.4091}
}

@article{TSMahesh_2022,
  title = {Experimental investigation of a quantum battery using star-topology NMR spin systems},
  author = {Joshi, Jitendra and Mahesh, T. S.},
  journal = {Phys. Rev. A},
  volume = {106},
  issue = {4},
  pages = {042601},
  numpages = {8},
  year = {2022},
  month = {Oct},
  publisher = {American Physical Society},
  doi = {10.1103/PhysRevA.106.042601},
  url = {https://link.aps.org/doi/10.1103/PhysRevA.106.042601}
}

@article{Preskill2000,
  title = {Quantum information and physics: Some future directions},
  volume = {47},
  ISSN = {1362-3044},
  url = {http://dx.doi.org/10.1080/09500340008244031},
  DOI = {10.1080/09500340008244031},
  number = {2–3},
  journal = {Journal of Modern Optics},
  publisher = {Informa UK Limited},
  author = {Preskill,  John},
  year = {2000},
  month = feb,
  pages = {127–137}
}

@book{Dutta2015, 
place={Cambridge}, 
title={Quantum Phase Transitions in Transverse Field Spin Models: From Statistical Physics to Quantum Information}, 
publisher={Cambridge University Press}, 
author={Dutta, Amit and Aeppli, Gabriel and Chakrabarti, Bikas K. and Divakaran, Uma and Rosenbaum, Thomas F. and Sen, Diptiman}, year={2015}
}

@article{Vidal_QPT,
  title = {Entanglement in Quantum Critical Phenomena},
  author = {Vidal, G. and Latorre, J. I. and Rico, E. and Kitaev, A.},
  journal = {Phys. Rev. Lett.},
  volume = {90},
  issue = {22},
  pages = {227902},
  numpages = {4},
  year = {2003},
  month = {Jun},
  publisher = {American Physical Society},
  doi = {10.1103/PhysRevLett.90.227902},
  url = {https://link.aps.org/doi/10.1103/PhysRevLett.90.227902}
}

@Article{Osterloh2002,
author={Osterloh, A.
and Amico, Luigi
and Falci, G.
and Fazio, Rosario},
title={Scaling of entanglement close to a quantum phase transition},
journal={Nature},
year={2002},
month={Apr},
day={01},
volume={416},
number={6881},
pages={608-610},
issn={1476-4687},
doi={10.1038/416608a},
url={https://doi.org/10.1038/416608a}
}

@article{Zhang_2007,
  title = {Disentanglement of two qubits coupled to an $XY$ spin chain: Role of quantum phase transition},
  author = {Yuan, Zi-Gang and Zhang, Ping and Li, Shu-Shen},
  journal = {Phys. Rev. A},
  volume = {76},
  issue = {4},
  pages = {042118},
  numpages = {7},
  year = {2007},
  month = {Oct},
  publisher = {American Physical Society},
  doi = {10.1103/PhysRevA.76.042118},
  url = {https://link.aps.org/doi/10.1103/PhysRevA.76.042118}
}

@article{DM_battery,
author = {Zhang, Xiang-Long and Song, Xue-Ke and Wang, Dong},
title = {Quantum Battery in the Heisenberg Spin Chain Models with Dzyaloshinskii-Moriya Interaction},
journal = {Advanced Quantum Technologies},
volume = {7},
number = {9},
pages = {2400114},
doi = {https://doi.org/10.1002/qute.202400114},
url = {https://advanced.onlinelibrary.wiley.com/doi/abs/10.1002/qute.202400114},
year = {2024}
}

@article{DM_battery_2,
doi = {10.1088/1402-4896/ad95c5},
url = {https://doi.org/10.1088/1402-4896/ad95c5},
year = {2024},
month = {dec},
publisher = {IOP Publishing},
volume = {100},
number = {1},
pages = {015106},
author = {K, Sanah Rahman and Murugesh, S},
title = {Effect of DM Interaction in the charging process of a Heisenberg spin chain quantum battery},
journal = {Physica Scripta}
}

@article{Banerjee_2007,
author={Banerjee, S.
and Ravishankar, V.
and Srikanth, R.},
title={Entanglement dynamics in two-qubit open system interacting with asqueezed thermal bath via quantum nondemolition interaction},
journal={The European Physical Journal D},
year={2010},
month={Jan},
day={01},
volume={56},
number={2},
pages={277-290},
issn={1434-6079},
doi={10.1140/epjd/e2009-00286-2},
url={https://doi.org/10.1140/epjd/e2009-00286-2}
}

@article{FICEK2002,
title = {Entangled states and collective nonclassical effects in two-atom systems},
journal = {Physics Reports},
volume = {372},
number = {5},
pages = {369-443},
year = {2002},
issn = {0370-1573},
doi = {https://doi.org/10.1016/S0370-1573(02)00368-X},
url = {https://www.sciencedirect.com/science/article/pii/S037015730200368X},
author = {Z. Ficek and R. Tanaś}
}

@misc{bhattacharya2025,
      title={Heisenberg spin chain models for realising quantum battery with the aid of Dzyaloshinskii Moriya interaction}, 
      author={Suprabha Bhattacharya and Vivek Balasaheb Sabale and Atul Kumar},
      year={2025},
      eprint={2508.20529},
      archivePrefix={arXiv},
      primaryClass={quant-ph},
      url={https://arxiv.org/abs/2508.20529}, 
}

@article{Uwe_Fischer_QB,
    author = {Gyhm, Ju-Yeon and Fischer, Uwe R.},
    title = {Beneficial and detrimental entanglement for quantum battery charging},
    journal = {AVS Quantum Science},
    volume = {6},
    number = {1},
    pages = {012001},
    year = {2024},
    month = {01},
    issn = {2639-0213},
    doi = {10.1116/5.0184903},
    url = {https://doi.org/10.1116/5.0184903}
}

@article{topological_QBs,
  title = {Topological Quantum Batteries},
  author = {Lu, Zhi-Guang and Tian, Guoqing and L\"u, Xin-You and Shang, Cheng},
  journal = {Phys. Rev. Lett.},
  volume = {134},
  issue = {18},
  pages = {180401},
  numpages = {8},
  year = {2025},
  month = {May},
  publisher = {American Physical Society},
  doi = {10.1103/PhysRevLett.134.180401},
  url = {https://link.aps.org/doi/10.1103/PhysRevLett.134.180401}
}

@misc{kading2025,
      title={Density matrices in quantum field theory: Non-Markovianity, path integrals and master equations}, 
      author={Christian Käding and Mario Pitschmann},
      year={2025},
      eprint={2503.08567},
      archivePrefix={arXiv},
      primaryClass={hep-th},
      url={https://arxiv.org/abs/2503.08567}, 
}

@article{rishav_QB,
  title = {Two-Time Weak-Measurement Protocol for Ergotropy Protection in Open Quantum Batteries},
  author = {Malavazi, Andr\'e H.A. and Sagar, Rishav and Ahmadi, Borhan and Dieguez, Pedro R.},
  journal = {PRX Energy},
  volume = {4},
  issue = {2},
  pages = {023011},
  numpages = {28},
  year = {2025},
  month = {Jun},
  publisher = {American Physical Society},
  doi = {10.1103/bv4w-jr6q},
  url = {https://link.aps.org/doi/10.1103/bv4w-jr6q}
}

@misc{malavazi2025_QB,
      title={Charge-Preserving Operations in Quantum Batteries}, 
      author={André H. A. Malavazi and Borhan Ahmadi and Paweł Horodecki and Pedro R. Dieguez},
      year={2025},
      eprint={2510.25549},
      archivePrefix={arXiv},
      primaryClass={quant-ph},
      url={https://arxiv.org/abs/2510.25549}, 
}

@misc{mukherjee2024_QB,
      title={Enhancement of an Unruh-DeWitt battery performance through quadratic environmental coupling}, 
      author={Arnab Mukherjee and Sunandan Gangopadhyay and A. S. Majumdar},
      year={2024},
      eprint={2411.02849},
      archivePrefix={arXiv},
      primaryClass={gr-qc},
      url={https://arxiv.org/abs/2411.02849}, 
}

@article{AhmadiB_2024nonreciprocal,
  title = {Nonreciprocal Quantum Batteries},
  author = {Ahmadi, B. and Mazurek, P. and Horodecki, P. and Barzanjeh, S.},
  journal = {Phys. Rev. Lett.},
  volume = {132},
  issue = {21},
  pages = {210402},
  numpages = {7},
  year = {2024},
  month = {May},
  publisher = {American Physical Society},
  doi = {10.1103/PhysRevLett.132.210402},
  url = {https://link.aps.org/doi/10.1103/PhysRevLett.132.210402}
}

@article{Ahmadi_2025superoptimal,
  title = {Superoptimal charging of quantum batteries via reservoir engineering: Arbitrary energy transfer unlocked},
  author = {Ahmadi, Borhan and Mazurek, Pawe\l{} and Barzanjeh, Shabir and Horodecki, Pawe\l{}},
  journal = {Phys. Rev. Appl.},
  volume = {23},
  issue = {2},
  pages = {024010},
  numpages = {14},
  year = {2025},
  month = {Feb},
  publisher = {American Physical Society},
  doi = {10.1103/PhysRevApplied.23.024010},
  url = {https://link.aps.org/doi/10.1103/PhysRevApplied.23.024010}
}

@article{ahmadi2025harnessing,
  title={Harnessing environmental noise for quantum energy storage},
  author={Ahmadi, Borhan and Ravichandran, Aravinth Balaji and Mazurek, Pawe{\l} and Barzanjeh, Shabir and Horodecki, Pawe{\l}},
  journal={arXiv preprint arXiv:2510.06384},
  year={2025}
}

@article{Albert_2023,
author = {Fert ,Albert and Chshiev ,Mairbek and Thiaville ,Andr\'{e} and Yang ,Hongxin},
title = {From Early Theories of Dzyaloshinskii–Moriya Interactions in Metallic Systems to Today’s Novel Roads},
journal = {Journal of the Physical Society of Japan},
volume = {92},
number = {8},
pages = {081001},
year = {2023},
doi = {10.7566/JPSJ.92.081001},
URL = {https://doi.org/10.7566/JPSJ.92.081001}
}

@article{MulaB_2023,
  title = {Ergotropy and entanglement in critical spin chains},
  author = {Mula, Bego\~na and Fern\'andez, Eva M. and Alvarellos, Jos\'e E. and Fern\'andez, Julio J. and Garc\'{\i}a-Aldea, David and Santalla, Silvia N. and Rodr\'{\i}guez-Laguna, Javier},
  journal = {Phys. Rev. B},
  volume = {107},
  issue = {7},
  pages = {075116},
  numpages = {8},
  year = {2023},
  month = {Feb},
  publisher = {American Physical Society},
  doi = {10.1103/PhysRevB.107.075116},
  url = {https://link.aps.org/doi/10.1103/PhysRevB.107.075116}
}

@article{parkavi2026tunable,
doi = {10.1088/1402-4896/ae30b0},
url = {https://doi.org/10.1088/1402-4896/ae30b0},
year = {2026},
month = {jan},
publisher = {IOP Publishing},
volume = {101},
number = {1},
pages = {015102},
author = {Parkavi, J Ramya and Muthuganesan, R and Chandrasekar, V K},
title = {Tunable dynamics of a dipolar quantum battery: role of spin-spin interactions and coherence},
journal = {Physica Scripta}
}

@article{vigneshwar2026noise,
doi = {10.1088/1751-8121/ae30b8},
url = {https://doi.org/10.1088/1751-8121/ae30b8},
year = {2026},
month = {jan},
publisher = {IOP Publishing},
volume = {59},
number = {1},
pages = {015302},
author = {Vigneshwar, B and Sankaranarayanan, R},
title = {Noise resilience of spin quantum battery in the presence of DM interactions},
journal = {Journal of Physics A: Mathematical and Theoretical}
}

\end{document}